\documentclass{article}
\usepackage{hyperref}
\hypersetup{
colorlinks = true, 
linkcolor=black, 
citecolor=black, 
urlcolor=black 
}
\usepackage{amsmath,amssymb,amsfonts}
\usepackage{algorithmic}
\usepackage{booktabs}
\usepackage{graphicx}
\usepackage{textcomp}
\usepackage{xcolor}
\usepackage{paralist}
\usepackage{cleveref}
\usepackage{subcaption}
\usepackage[most]{tcolorbox}
\usepackage{multirow}

\usepackage{bbm}
\usepackage{paralist}

\newcommand{\inlineimage}[1]{%
\raisebox{-.2\height}{\includegraphics[width=1em]{#1}}%
}
\newcommand{\inlineimageb}[1]{%
\raisebox{-.3\height}{\includegraphics[width=0.7em]{#1}}%
}

\newcommand{\Atwelve}{\ensuremath{\hat{A}}\textsubscript{12}\xspace}

\usepackage{hyperref}
\hypersetup{
colorlinks = true, 
linkcolor=black, 
citecolor=black, 
urlcolor=black 
}

\usepackage{amsfonts} 
\usepackage{nicefrac} 

\usepackage{microtype} 

\usepackage{import}
\usepackage{float}
\usepackage{multirow}
\usepackage{paralist}
\usepackage{arydshln}
\usepackage{stfloats}
\usepackage{xspace}
\usepackage{braket}
\usepackage{lscape}

\usepackage{PRIMEarxiv}

\title{Assessing Quantum Extreme Learning Machines for Software Testing in Practice}

\author{
    Asmar~Muqeet\\
    Simula Research Laboratory \\
    University of Oslo \\
    Oslo\\
    \texttt{asmar@simula.no} \\
    \And
    Hassan Sartaj\\
    Simula Research Laboratory\\
    Oslo, Norway \\
    \texttt{hassan@simula.no}\\
    \And
    Aitor Arrieta\\
    Mondragon University\\
    Mondragon, Spain \\
    \texttt{aarrieta@mondragon.edu}\\
    \And
    Shaukat~Ali \\
    Simula Research Laboratory and \\
    Oslo Metropolitan University \\
    Oslo\\
    \texttt{shaukat@simula.no} \\
    \And
    Paolo~Arcaini \\
    National Institute of Informatics \\
    Tokyo\\
    \texttt{arcaini@nii.ac.jp} \\
    \And
    Maite Arratibel\\
    Orona\\
    San Sebastian, Spain \\
    \texttt{marratibel@orona-group.com}\\
    \And
    Julie Marie Gjøby\\
    Welfare Technologies Section,\\
    Oslo Kommune Helseetaten\\
    Oslo, Norway \\
    \texttt{julie-marie.gjoby@hel.oslo.kommune.no}\\
    \And
    Narasimha Raghavan Veeraragavan\\
    Cancer Registry of Norway,\\
    Norwegian Institute of Public Health,\\ 
    Oslo, Norway\\
    \texttt{nara@kreftregisteret.no}\\
    \And
    Jan~F.~Nygård\\
    Cancer Registry of Norway,\\
    Oslo, Norway\\
    and The Arctic University of Norway\\
    Tromsø, Norway \\
    \texttt{jfn@kreftregisteret.no}
}

\begin{document}
\maketitle

\begin{abstract}
Machine learning has been extensively applied for classical software testing activities such as test generation, minimization, and prioritization. Along the same lines, there has been interest in applying quantum machine learning to classical software testing. For example, Quantum Extreme Learning Machines~(QELMs) were recently applied for testing classical software of industrial elevators. However, studies on QELMs, whether in software testing or other areas, used ideal simulators that fail to account for the noise in current quantum computers. While ideal simulations offer insight into QELM's theoretical capabilities, they do not enable studying their performance on current noisy quantum computers. To this end, we study how quantum noise affects QELM in three industrial classical software testing case studies, providing insights into QELMs' robustness to noise for software testing applications. Such insights assess QELMs potential as a viable solution for software testing problems in today's noisy quantum computing. Our results show that QELMs are significantly affected by quantum noise, with a performance drop of 250\% in regression and 50\% in classification software testing tasks. Quantum noise also increases uncertainty in QELM models, producing a saturation effect where larger qubit counts make the models increasingly random and unreliable. While error mitigation techniques can enhance noise resilience—achieving an average 3\% performance drop in classification—their effectiveness varies by context. For classification tasks, QLEAR performs well, whereas Zero Noise Extrapolation is more effective for regression and smaller qubit counts. However, no single mitigation approach consistently reduces uncertainty across tasks or scales reliably as the number of qubits increases, highlighting the need for QELM-tailored strategies.
\end{abstract}

\keywords{Software Testing, Quantum Computing, Machine learning, Quantum Noise, Quantum Machine Learning, Quantum circuit, Uncertainty Quantification}

\section{Introduction}
Machine learning~(ML) has enhanced software testing of classical software systems for various activities~\cite{mlforse, Kotti2023}. Along the same lines, quantum machine learning~(QML)~\cite{qml} is gaining interest in this field~\cite{xinyiqaoa,quell}, aiming to further improve machine learning-assisted software testing with Quantum Computing (QC), which uses quantum-mechanical principles to perform computational tasks~\cite{qc}. Consequently, it is expected to provide exponential speedups for ML algorithms. Within QML, Quantum Extreme Learning Machines (QELMs) offer an innovative approach to enhance information processing capabilities, leading to efficient training of classical linear ML models that can match or even outperform complex ML models~\cite{quell,qelmapplication}.

QELMs can potentially address several challenges in machine learning-based software testing, such as high training data cost, test automation complexity, and test environment variability~\cite{softwaretestingchallenge}. This is possible due to their ability to map data into higher-dimensional quantum states, which facilitates the development of simpler and more generalizable models~\cite{binary_class}. A recent study demonstrated the effectiveness of QELMs in regression testing for industrial elevator software by developing a machine learning model that requires fewer features for predictions compared to the classical machine learning model~\cite{quell}. However, most studies on QELMs used ideal quantum simulators that do not incorporate quantum noise. While ideal simulations provide valuable insights into the theoretical capabilities of QELMs, they do not reflect the reality, since current quantum computers are noisy. Such noise significantly impacts computational accuracy. On the other hand, ideal quantum simulators have high computational costs~\cite{simulators_are_slow}, and simulating large-scale problems with classical computers is often unfeasible~\cite{simulationlimit}. Consequently, the effectiveness of QELMs in real-world applications of classical software testing, especially in the presence of quantum noise, remains largely unexplored.

This paper bridges the gap between ideal simulations and real-world noisy quantum computations by examining the impact of quantum noise on QELM models through the following 
case studies of classical software testing:
\begin{enumerate}
\item {\it Industrial Elevator Software}: Orona\footnote{\url{https://www.orona-group.com/int-en/}}--a world leader in building elevators provides this case study. Their dataset is used for machine learning (ML)-based regression testing of industrial elevator software~\cite{quell}; 
\item {\it IoT Application Testing with Medical Device Digital Twins}: The second case study is from Oslo City's Health Care department, which is responsible for providing healthcare services to its residents, including advanced home care with specialized medical devices. We used a dataset from one particular medical device, i.e., the Karie medicine dispenser~\cite{karie}. This dataset is used for testing an IoT-based healthcare application with ML-based digital twins of medical devices~\cite{sartaj2023testing,sartaj2024medet};
\item {\it Testing Cancer Registration and Support System (CaReSS)}: The third case study is provided by the Norwegian Cancer Registry's CaReSS software system. The dataset from CaReSS is used for ML-based cost-effective testing~\cite{isaku2023cost}.
\end{enumerate}

We explore how quantum noise impacts the accuracy and reliability of QELMs under real-world conditions for software testing problems. To cope with quantum noise, several error mitigation methods have been proposed~\cite{mitiq,qlear,qoin,quietIEEESoftware2025}. Therefore, we also assess the feasibility of combining QELMs with noise error mitigation techniques to enhance their applicability on current quantum computers. For reliability analysis, accuracy alone is not sufficient under quantum noise; therefore, we also incorporated uncertainty quantification to study how quantum noise impacts the uncertainty of QELM models. By focusing on practical applications, we aim to better understand QELMs' resilience to quantum noise and determine their potential as viable solutions for challenges in today's noisy quantum computing era.

To summarize, our main contributions are as follows:
\begin{enumerate}
\item We empirically evaluate the resistance of QELM models to quantum noise through three real-world software testing case studies from the current practice of one commercial company from Spain and two public sectors from Norway.
\item We assess the performance of QELM models on three real quantum computers from IBM, utilizing their noise models, thereby making it one of the first studies to assess QELMs in noisy conditions for classical software testing. 
\item We examine the applicability of quantum noise mitigation techniques to QELMs in the current era of noisy quantum computers.
\item We study uncertainty quantification for QELMs in real-world use cases, establishing a baseline for reliability analysis under quantum noise.
\end{enumerate}

Our results show that QELMs are highly impacted by quantum noise. In the Orona dataset, QELM performance dropped of 250\% across all noise models during the ML testing phase, highlighting poor results in the regression task. Incorporating noise into both the ML training and testing pipeline improved performance to just above 50\%, but the deviation from ideal remained significant. For classification tasks, performance decreased by 50\% in the Karie and CaReSS datasets. Noise in the ML pipeline improved CaReSS performance to 30\%, but no improvement for the Karie dataset. Furthermore, we observed that, in the context of software testing, ML-based error mitigation methods perform well in classification tasks, but their effectiveness depends on the context. ML-based methods performed well in classification, with only a 3\% deviation from ideal when introduced in both the ML training and testing phase, but faced challenges in regression tasks. 

In terms of uncertainty, quantum noise significantly increases uncertainty in QELM models, producing a saturation effect, i.e., predictive uncertainty in QELMs rises due to quantum noise, but as the qubit count increases, it eventually levels off at a high value. This saturation effect occurs because the accumulated quantum noise overwhelms the useful information, leaving the model in a state of persistent unreliability. While error mitigation can partially reduce the uncertainty, its effectiveness depends on the task and method. The QLEAR noise mitigation method shows strong performance for classification, whereas the ZNE method is more effective for regression and smaller qubit counts. Importantly, no single mitigation approach consistently controls uncertainty across all tasks or scales reliably with qubit number. This underscores that the practical application of QELMs in the current noisy quantum computing era is restricted, encouraging QELM-tailored error mitigation strategies to improve their performance for practical use in software engineering problems and beyond.

\paragraph{Paper structure: }
Section~\ref{sec:background} introduces background notions necessary for understanding the paper, and Section~\ref{sec:applicationContext} presents the three real-world case studies that we use to conduct the study. Section~\ref{sec:experimentDesign} describes the design of the experiments, and Section~\ref{sec:experimentalResults} presents the experimental results. Section~\ref{sec:threats} identifies the threats that may affect the validity of the study. Section~\ref{sec:discussion} provides a more detailed discussion about the results and the lessons learnt. Finally, Section~\ref{sec:related} reviews related work, and Section~\ref{sec:conclusion} concludes the paper.

\section{Background}\label{sec:background}

\subsection{Quantum Computing Basics}
This section provides a brief overview of the fundamental concepts of quantum computing and quantum noise.

\textbf{Qubits:} In classical computing, information is represented using bits, which can only be in one of two states: 0 or 1. In quantum computing~(QC), however, the basic unit of information is the quantum bit, or \textit{qubit}, which can exist in a \textit{superposition} of both $|0\rangle$ and $|1\rangle$ states simultaneously, each with a corresponding amplitude $(\alpha)$. This amplitude is a complex number, characterized by a \textit{magnitude} and a \textit{phase} when represented in polar form. The state of a qubit is expressed using the Dirac notation~\cite{dirac} as: $|\psi\rangle = \alpha_0 |0\rangle + \alpha_1 |1\rangle$, where $\alpha_0$ and $\alpha_1$ are the amplitudes associated with the $|0\rangle$ and $|1\rangle$ states, respectively. The likelihood of measuring the qubit in either state is given by the square of the magnitude of its amplitude, with the total probability always summing to 1: $|\alpha_0|^2 + |\alpha_1|^2 = 1$.

\begin{figure}[!tb]
\centering
\includegraphics[width=0.9\textwidth]{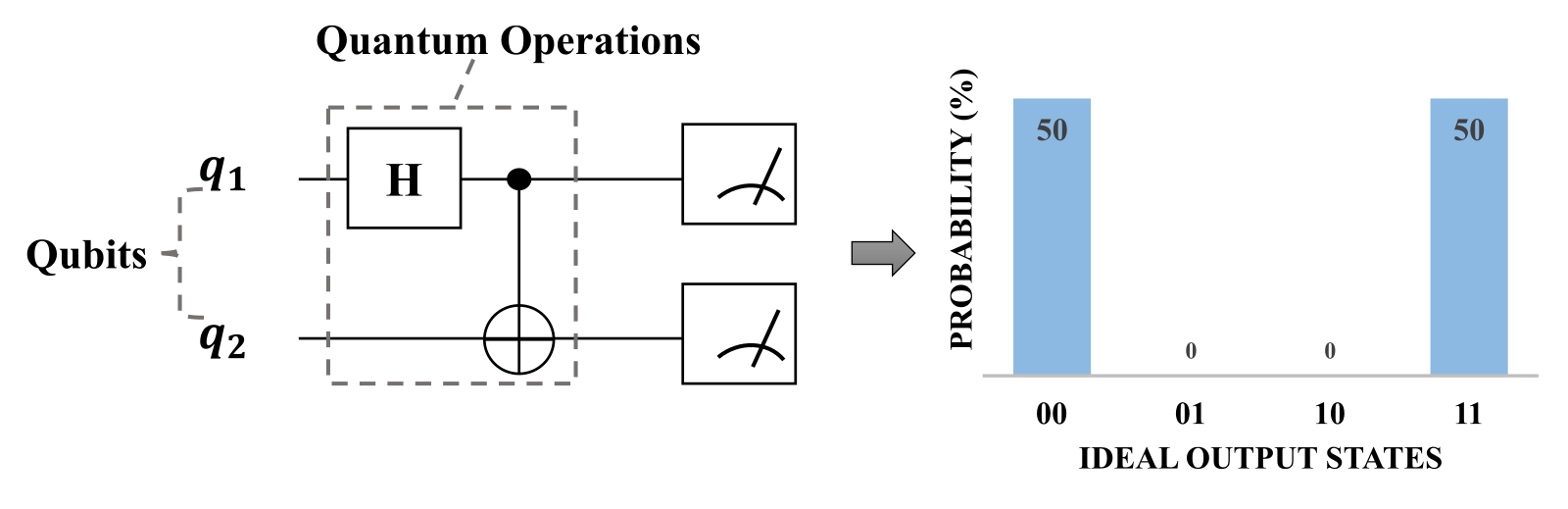}
\caption{A Two-qubit quantum circuit that creates entanglement between two qubits. The circuit output is an equal probability distribution of two entangled states.}
\label{fig:twoqubit}
\end{figure}

\textbf{Quantum Circuits:}
Quantum computers are currently programmed using quantum circuits, which consist of a sequence of quantum gates that manipulate the states of qubits. A simple example of a quantum gate is the \textit{NOT} gate, which flips a qubit's state from $|0\rangle$ to $|1\rangle$ or vice versa. In quantum circuits, various quantum gates are applied to create superposition and entanglement between qubits, both of which are essential for processing information. Entanglement is a unique quantum phenomenon where the state of one qubit is directly correlated with the state of another. Figure~\ref{fig:twoqubit} presents a two-qubit quantum circuit and its corresponding output.
This circuit is designed to entangle two qubits. Initially, both qubits ($q_1$ and $q_2$) are in the state $|0\rangle$. First, a \textit{Hadamard} gate~\inlineimage{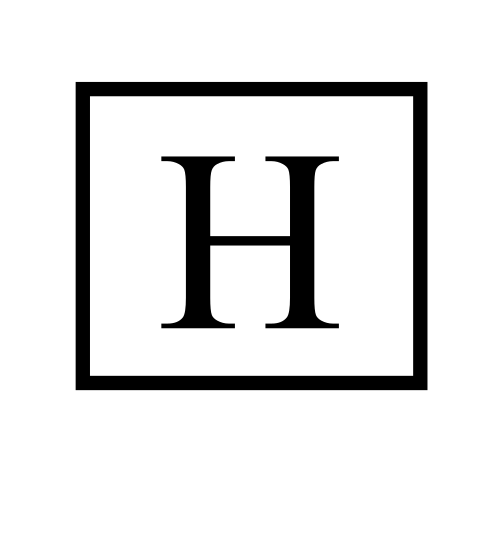} is applied to $q_1$, putting it into a superposition of $|0\rangle$ and $|1\rangle$. Then, a \textit{controlled-NOT} gate~\inlineimageb{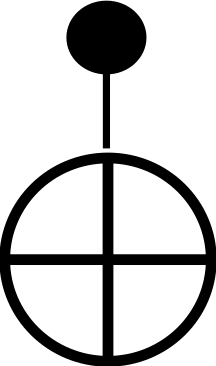} is used to entangle $q_1$ with $q_2$. After these operations, a measurement operation~\inlineimage{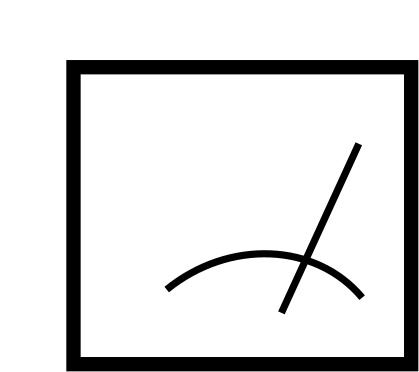} is performed to collapse the qubits' superposition and entanglement into a definite state of either $|0\rangle$ or $|1\rangle$. The quantum circuit is executed multiple times to generate a probability distribution of the results. This output consists of binary strings, representing the qubits' final states, and their corresponding probabilities indicate how often each outcome occurs over repeated runs.

\subsection{Quantum Noise}
\begin{figure}[!tb]
\centering
\includegraphics[width=0.8\textwidth]{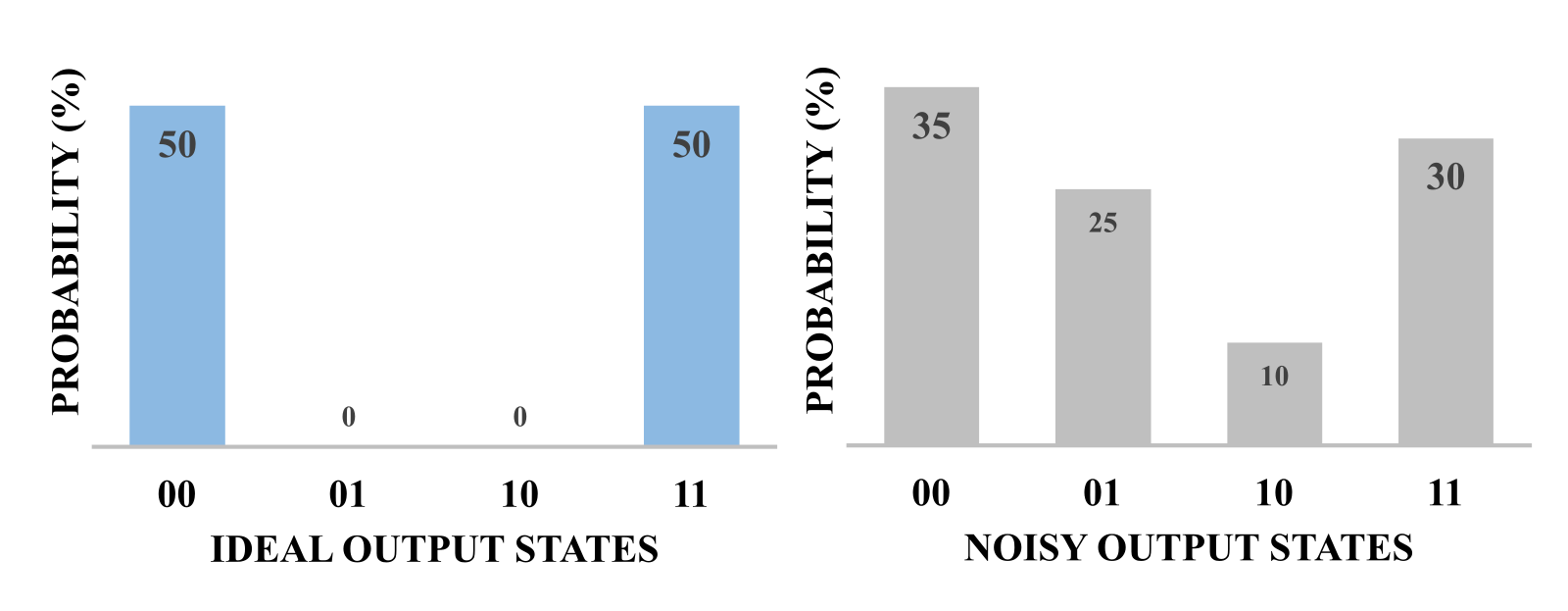}
\caption{Two-qubit entanglement quantum circuit: ideal output and noisy output.}
\label{fig:idealvsnoise}
\end{figure}

Current quantum computers are susceptible to quantum noise, which compromises the precision of their computations. Quantum noise arises from various sources. First, environmental factors such as magnetic fields and radiation can impact quantum operations~\cite{noise_benchmark1}. Interactions between qubits and their environments can lead to disturbances and loss of information in quantum states, a phenomenon known as \emph{decoherence}~\cite{decoherence_def}. Second, unwanted interactions between qubits, even when perfectly isolated from their surroundings, can produce \emph{crosstalk noise}~\cite{crosstalkgatenoise}, which leads to unintended quantum states. Third, inaccuracies in hardware calibration for quantum gate operations also contribute to noise. Minor calibration errors can result in slight changes in phase or amplitude in qubits. While these changes may not immediately destroy a quantum state, they can lead to undesirable states following a series of gate operations~\cite{crosstalkgatenoise}. It is important to note that any qubit in a circuit can be influenced by noise at any stage, resulting in an accumulated effect on the circuit's output. Figure~\ref{fig:idealvsnoise} illustrates the difference between the ideal and noisy outputs of the entanglement circuit shown in Figure~\ref{fig:twoqubit}.
In the noisy output, states 01 and 10 emerge due to quantum noise, which alters the overall probability distribution of the two entangled qubits compared to the ideal output.

Due to quantum noise in real quantum computers, researchers frequently rely on quantum simulators to run circuits without noise interference~\cite{simulators}. However, ideal quantum simulators are limited by classical computing resources and quickly become impractical, even when using supercomputers, as the number of qubits and operations in a circuit increases~\cite{simulationlimit}. As a result, they are only suitable for circuits with a small number of qubits. In addition to ideal simulators, QC platforms like IBM and Google offer noise models that simulate the noise experienced by their quantum computers. These \textit{\textbf{noise models}} replicate various types of noise, such as decoherence, gate errors, and crosstalk, helping researchers understand how noise impacts quantum computations and enabling error mitigation strategies. Despite their usefulness, noisy simulations also face the same computational limitations as ideal simulators and are only used to approximate real quantum computer behavior for circuits with a small number of qubits.

\subsection{Quantum Extreme Learning Machine}\label{sec:qelm}
Extreme Learning Machine (ELM) is an algorithm proposed to address the limitations of gradient-based backpropagation neural networks~\cite{elmsurvey}. ELM is a single-layer or multi-layer feedforward network where hidden layer weights are randomly assigned and fixed, eliminating the need for iterative tuning of weights via backpropagation. Instead of using gradient descent to minimize an error function, the output of the neural network is used to train a simple linear model for prediction. This approach enables ELM to achieve fast training speeds while maintaining good generalization performance~\cite{elmsurvey}. Quantum Extreme Learning Machine (QELM) is the quantum counterpart of the classical ELM. QELM integrates principles of quantum computing to further enhance the computational power and efficiency of ELMs~\cite{qelm}. 

\begin{figure}[!tb]
\centering
\includegraphics[width=0.99\columnwidth]{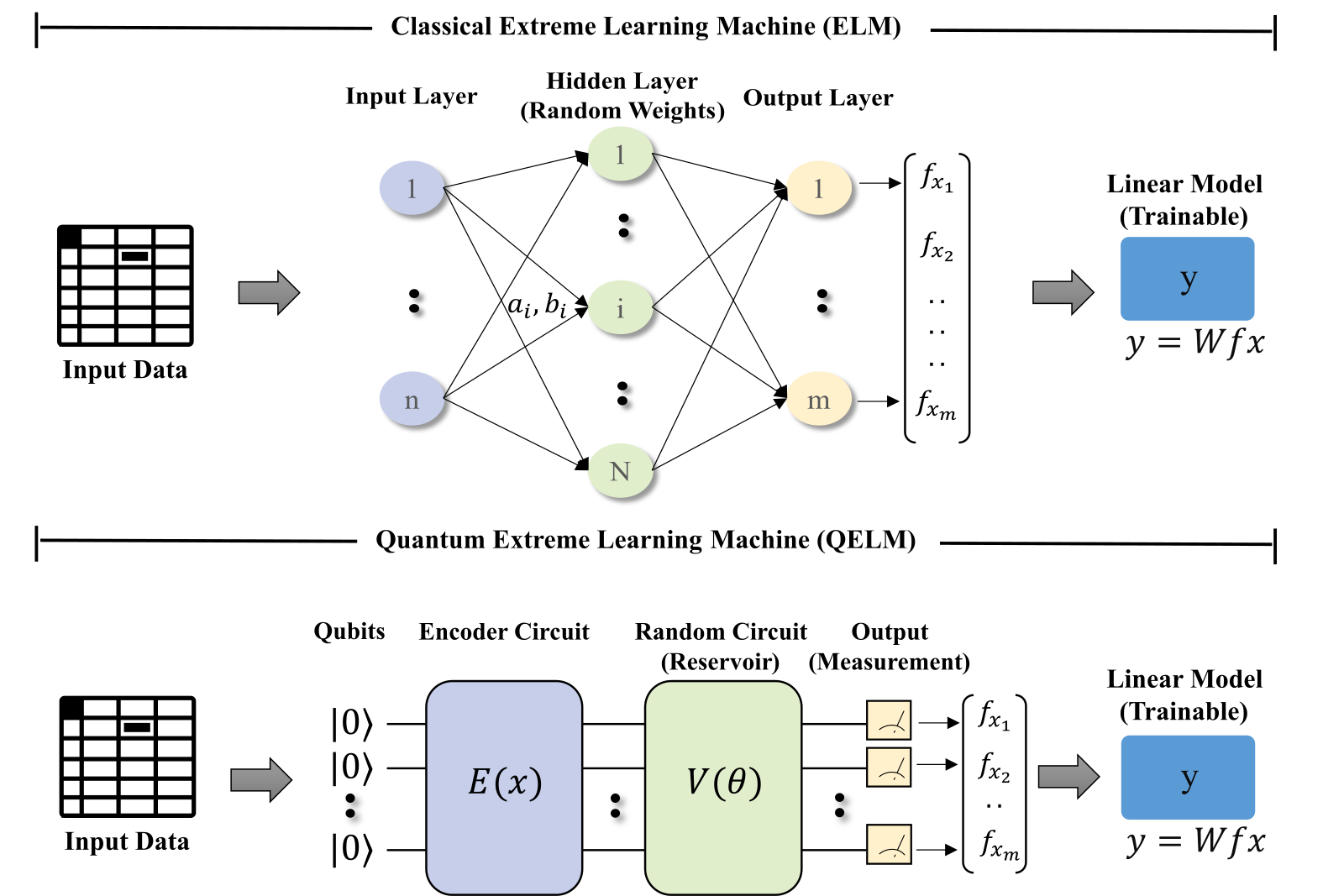}
\caption{Comparison between classical extreme learning machine and quantum extreme learning machine. The semantically equivalent components in both algorithms are highlighted using the same colors.}
\label{fig:elmvsqelm}
\end{figure}

Figure~\ref{fig:elmvsqelm} compares ELM and QELM, where semantically equivalent components are highlighted with the same color. 
In classical ELM, the neural network's input layer connects the input data with the hidden layers. It is responsible for transforming the input into a format compatible with the hidden layers (e.g., converting text to numeric representations). In QELM, the equivalent of the input layer is an \textbf{\textit{Encoder}}, which is a quantum circuit responsible for converting classical data into quantum states. Next, in ELM, the randomly initialized hidden layers are responsible for feature extraction and transforming the data into a higher-dimensional feature space, which can be used to train the linear model. In QELM, the equivalent of the hidden layer is the \textbf{\textit{Reservoir}}, a quantum circuit that performs feature extraction and transforms data into a quantum feature space. Finally, in ELM, the output layer extracts the features created by the hidden layer, resulting in a feature vector that is then used to train a linear model for final prediction. In QELM, the measurement operation on the reservoir circuit takes the role of the output layer, returning a classical feature vector after collapsing the superposition and entanglement of qubits.

\subsection{Uncertainty Quantification}\label{sec:uq}
Uncertainty Quantification (UQ) plays a critical role in evaluating the reliability and trustworthiness of predictive models. Indeed, conventional performance metrics such as accuracy, RMSE, or F1 score summarize overall performance but fail to capture how confident a model is in its individual predictions~\cite{UQbook}. UQ fills this gap by assessing the variability and stability of predictions, thereby providing a more nuanced understanding of model behavior.

In the case of QELMs, UQ becomes particularly important. Quantum noise can perturb output probability distributions and introduce instability into the decision-making process~\cite{crosstalkgatenoise}. As a result, a QELM may report high accuracy while still generating predictions with low confidence—something that traditional metrics cannot reveal but UQ can explicitly expose. By identifying cases in which predictions are less reliable, UQ ensures that decisions are informed not only by accuracy but also by the model's level of certainty. At present, no QELM-specific methods for UQ exist. For this reason, we focus on model-agnostic UQ techniques, which can be applied regardless of whether the underlying architecture is classical or quantum-inspired. Importantly, the choice of UQ methodology differs between regression and classification models.

\noindent\textbf{Uncertainty in Regression Models:}
In regression tasks, the goal of UQ is to characterize the range of plausible outcomes surrounding a point estimate. Since regression models produce continuous outputs, uncertainty is typically expressed in the form of intervals or variance estimates~\cite{UQbook}. The most widely used techniques include prediction intervals and scoring rules~\cite{uqmethods}. Prediction intervals define a range around the predicted mean that reflects the possible spread of the true outcome~\cite{uqmethods}. They serve as a direct and interpretable means of communicating uncertainty. For example, if a model predicts $y = 5.0$ with a 95\% prediction interval of $[4.2, 5.8]$, this indicates high confidence that the true outcome lies within that range. Scoring rules provide a quantitative framework that evaluates the trade-off between accuracy and reliability of the uncertainty estimates~\cite{uqmethods}. Both metrics require access to a distribution of predictions, which can be obtained using resampling and aggregation strategies. The most common strategies are Bootstrapping and Ensembling~\cite{UQbook}. Bootstrapping distribution is collected by repeatedly resampling the training dataset, fitting models to each resample, and then analyzing the variability across their predictions. Ensembling, on the other hand, constructs multiple independent models and aggregates their outputs. The diversity of predictions across ensemble models acts as a practical proxy for uncertainty, offering insight into the model's reliability beyond single-point estimates.

\noindent\textbf{Uncertainty in Classification Models:}
In classification settings, uncertainty is linked to the reliability of predicted class probabilities. Unlike regression, where uncertainty is commonly expressed through intervals, classification UQ focuses on the calibration and trustworthiness of probabilistic outputs~\cite{UQbook}. A well-calibrated model produces confidence estimates that align with empirical accuracy, ensuring that predicted probabilities carry meaningful interpretability. The most common methods from model agnostic classification UQ are Reliability Diagrams and Scoring rule metrics~\cite{uqmethods}. Reliability diagrams provide a graphical means of assessing calibration by comparing predicted confidence levels against observed accuracy. For a perfectly calibrated model, the curve lies along the diagonal: for instance, predictions made with 70\% confidence should indeed be correct 70\% of the time. Deviations from the diagonal reveal systematic biases such as overconfidence (probabilities higher than actual correctness) or underconfidence (probabilities lower than actual correctness). These diagrams thus serve as an intuitive tool for diagnosing whether classification probabilities are trustworthy. Scoring rules provide a quantitative framework that evaluates the trade-off between accuracy and reliability of the uncertainty estimates. Most widely used scoring rule metrics for classification are Brier Score and Log-Loss (Cross-Entropy)~\cite{uqmethods}.

\section{Application Context}\label{sec:applicationContext}

\begin{figure}[!tb]
\includegraphics[width=0.99\textwidth]{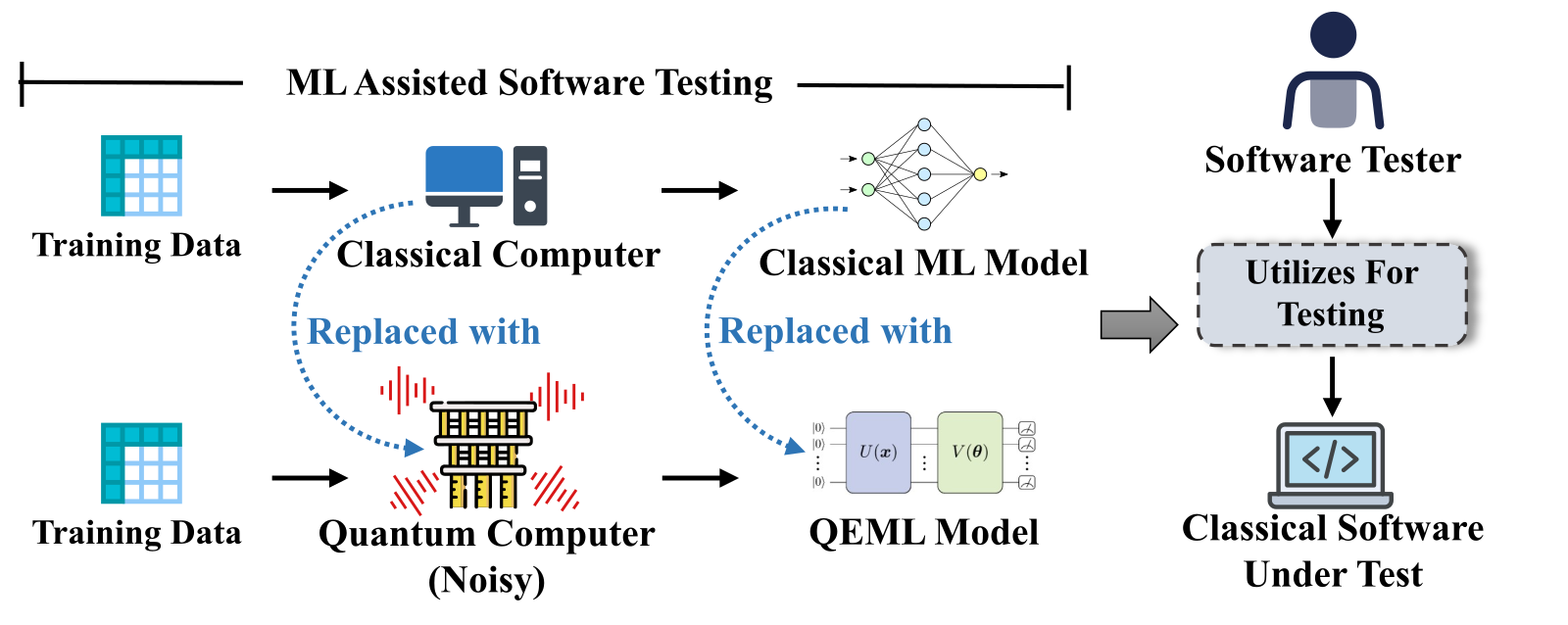}
\caption{Transition from classical ML-based software testing to QELM-based classical software testing}
\label{fig:testingfuture}
\end{figure}

Classical machine learning has been applied to support classical software testing~\cite{Kotti2023, mlforse}. In this context, quantum machine learning could replace classical machine learning by leveraging its advantages in computational speed, improved feature extraction, simplified model complexity, and enhanced overall performance~\cite{qmlbenifits} to support classical software testing. Along this line, a recent study assessed QELM for regression testing of classical software deployed on industrial elevators~\cite{quell}. The study highlighted the potential of QELM, where it reduced the number of machine learning features required with QELM compared to the ones required with classical ML models, demonstrating that QELM can outperform classical models in certain situations and has the potential to replace classical ML models for classical software testing. Figure~\ref{fig:testingfuture} illustrates our application context in which QELM is utilized instead of classical machine learning to support software testing.
In the figure, instead of using a classical machine learning model trained on classical data on a classical computer, the software tester employs a QELM model trained on the same classical training data utilizing a current noisy quantum computer. This QELM model can improve performance across various activities, such as regression testing, oracle estimation, and test optimization. Consequently, this setting replaces the classical computer and machine learning model with a quantum computer and QELM model to support the testing activities of the classical system under test.

In this study, we examine three distinct case studies in software testing. Each case study presents a real-world scenario where classical machine learning was applied to support classical software testing. Consequently, classical machine learning can be replaced with QELM to enhance the software testing of the classical software under test. Below, we define the application context for each case study.

\subsection{Orona Elevator}

\subsubsection{Application Context}

Orona is one of the largest elevator companies worldwide. Elevators aim to safely transport passengers in a building while trying to provide maximum comfort. To measure this comfort, different Quality of Service (QoS) metrics are employed. A well-known QoS metric in this domain is the Average Waiting Time (AWT) of passengers. The waiting time of a passenger refers to the time between the floor call until an elevator attends that call. AWT refers to the average time that a set of passengers need to wait until an elevator serves their call. This metric depends, to a large extent, on the software dispatching algorithm. This algorithm receives as input several data (e.g., the position of each elevator, the weight of each elevator, and their direction), and based on it, its goal is to determine, for each landing call, which is the elevator that should serve it.

\subsubsection{ML-based Regression Testing of Elevator Software}

\begin{figure}[!tb]
\centering
\includegraphics[width=0.9\textwidth]{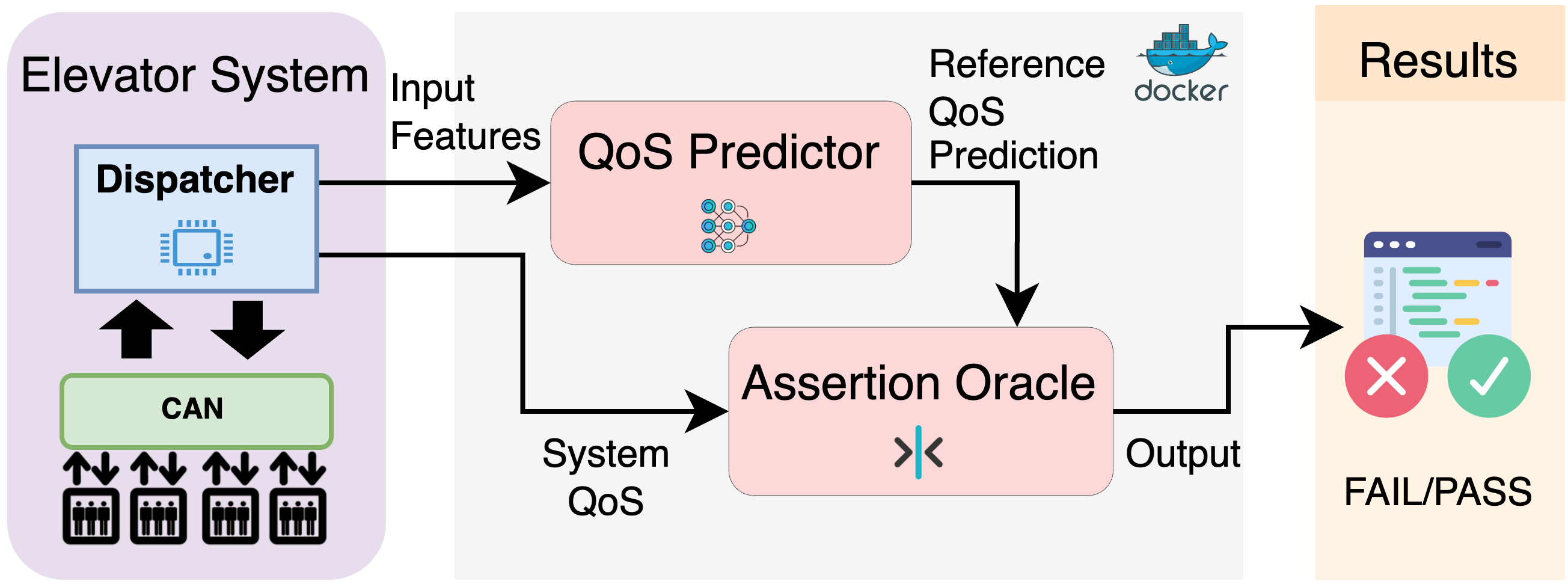}
\caption{An overview of Orona testing context}

\label{fig:orona}
\end{figure}

Orona has a suite of elevator dispatching algorithms implemented as software. As with any other software system, these elevator dispatching algorithms evolve over time to handle maintenance (e.g., bug corrections, inclusion of new functionalities, adaption to new hardware demands). Because such systems suffer from the test oracle problem~\cite{ayerdi2020qos}, the different versions are usually tested by regression test oracles. Given a set of passengers in a building, the AWTs provided by different dispatching algorithm versions are compared. Ideally, the new version should have better or similar AWT to the old version. These oracles, however, do not scale either when the software needs to be tested at the Hardware-in-the-Loop test levels or when deployed at operation (for run-time monitoring). Therefore, recently, Orona has explored the possibility of using machine learning to predict the QoS (e.g., the AWT) of elevators based on domain-specific inputs (e.g., number of active calls)~\cite{gartziandia2022machine}, substituting traditional regression test oracles (as shown in Figure~\ref{fig:orona}). 
Furthermore, a recent study showed that quantum extreme learning machines can outperform traditional machine-learning techniques in this problem~\cite{quell}. In the context of this paper, we will study whether the superior performance demonstrated by QELM in Orona's context in~\cite{quell} still holds under quantum noise. 

\subsection{Oslo City Healthcare Data}

\subsubsection{Application Context}The healthcare department of Oslo City~\cite{oslocity} aims to provide efficient and high-quality healthcare services to Oslo residents and extend these services to other counties in Norway. 
With this aim, Oslo City collaborates with different healthcare companies to develop an IoT-based platform that comprises an interconnected network of smart medical devices, third-party health services, pharmacies, medical professionals, health authorities, caregivers, and patients~\cite{sartaj2023hita}. 
In this platform, smart medical devices are essential in providing various health services~\cite{sartaj2024modelbased}. 
These devices are allocated to patients based on their healthcare needs and continuously alert key stakeholders, such as doctors and caregivers, especially in emergencies. 
For example, the Karie~\cite{karie} medicine dispenser is one of such types of smart medical devices, developed to deliver medications on time. 
It retrieves a medication plan from pharmacies, dispenses medicines at the specified time, generates reminder alarms for the patient, and notifies stakeholders about medication adherence. 
Karie also features a user-friendly interface that can be customized to individual patient preferences, either directly on the device or remotely by caregivers. 

\subsubsection{Testing with ML-based Digital Twins of Medical Devices}
\begin{figure}[!tb]
\centering
\includegraphics[width=0.8\textwidth]{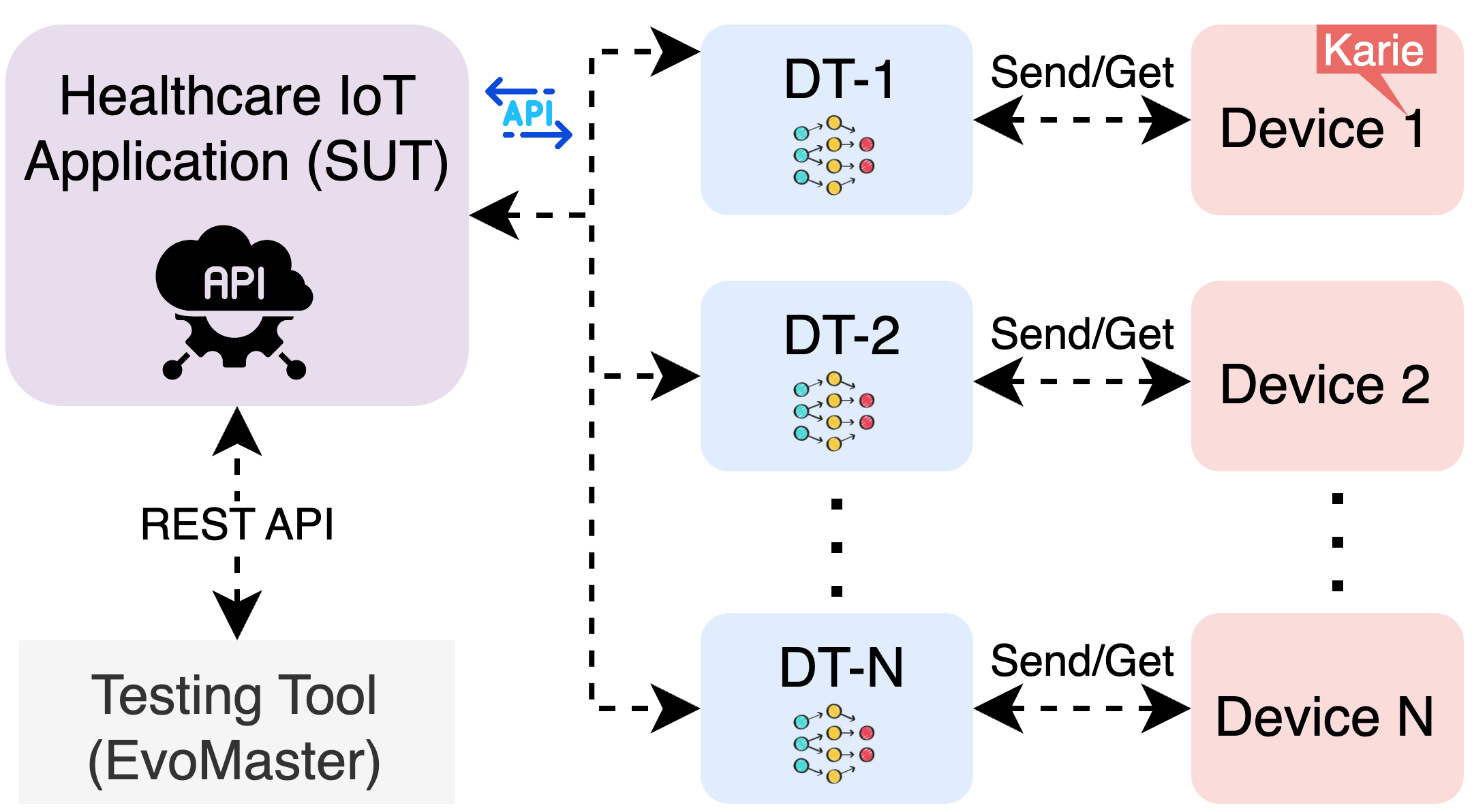}
\caption{An overview of testing an IoT-based healthcare application with DTs}
\label{fig:hiot}
\end{figure}

System and integration level testing of IoT-based healthcare applications requires incorporating multiple medical devices. 
However, employing these devices during test execution could potentially result in device damage or service blocking from device servers~\cite{sartaj2025restapi,sartaj2024uncertainty}. 
Given these challenges, we developed ML-based digital twins (DTs) of medical devices to facilitate automated and rigorous testing of IoT-based healthcare applications~\cite{sartaj2023hita}. 
Figure~\ref{fig:hiot} shows an overview of the test execution infrastructure with medical device DTs connected to the system under test (SUT).
In this setup, a testing tool like EvoMaster~\cite{arcuri2018evomaster} generates REST API tests for the SUT. 
During test execution, the SUT communicates with the DTs that manage all API calls. 
The DTs then communicate with the corresponding physical devices (e.g., Karie) for scenarios such as calibration or synchronization. In this study, we use the Karie device dataset as one of our case studies to assess the potential of QELM in the presence of quantum noise.

\subsection{Norway's Cancer Registry Data}

\subsubsection{Application Context} The Cancer Registry of Norway (CRN) collects data regarding cancer cases in Norway. This data is collected as {\it cancer messages} sent by various health institutes, such as laboratories and hospitals. A cancer message consists of all metadata related to cancer cases, such as diagnosis, treatment history, and follow-up details~\cite{isaku2023cost}. 
More specifically, each cancer message contains detailed information, including the cancer stage, type, topography, morphology, surgical interventions, and medical history. 
Cancer messages received from various sources are first validated using a well-defined set of rules and guidelines stipulated by international standards, such as ICD-10 and ICD-O-3. 
Subsequently, CRN analyzes cancer data and generates statistics to assist policymakers, healthcare authorities, and other relevant stakeholders in decision-making. 
For this purpose, CRN has developed an automated Cancer Registration Support System (CaReSS) to validate and analyze data, and generate statistics~\cite{laaber2023challenges}. 
The cancer data and statistics produced with CaReSS are made available to researchers to facilitate cancer research and enhance the quality of the cancer registry system.

\subsubsection{ML-based Testing of CaReSS}
\begin{figure}[!tb]
\centering
\includegraphics[width=0.8\textwidth]{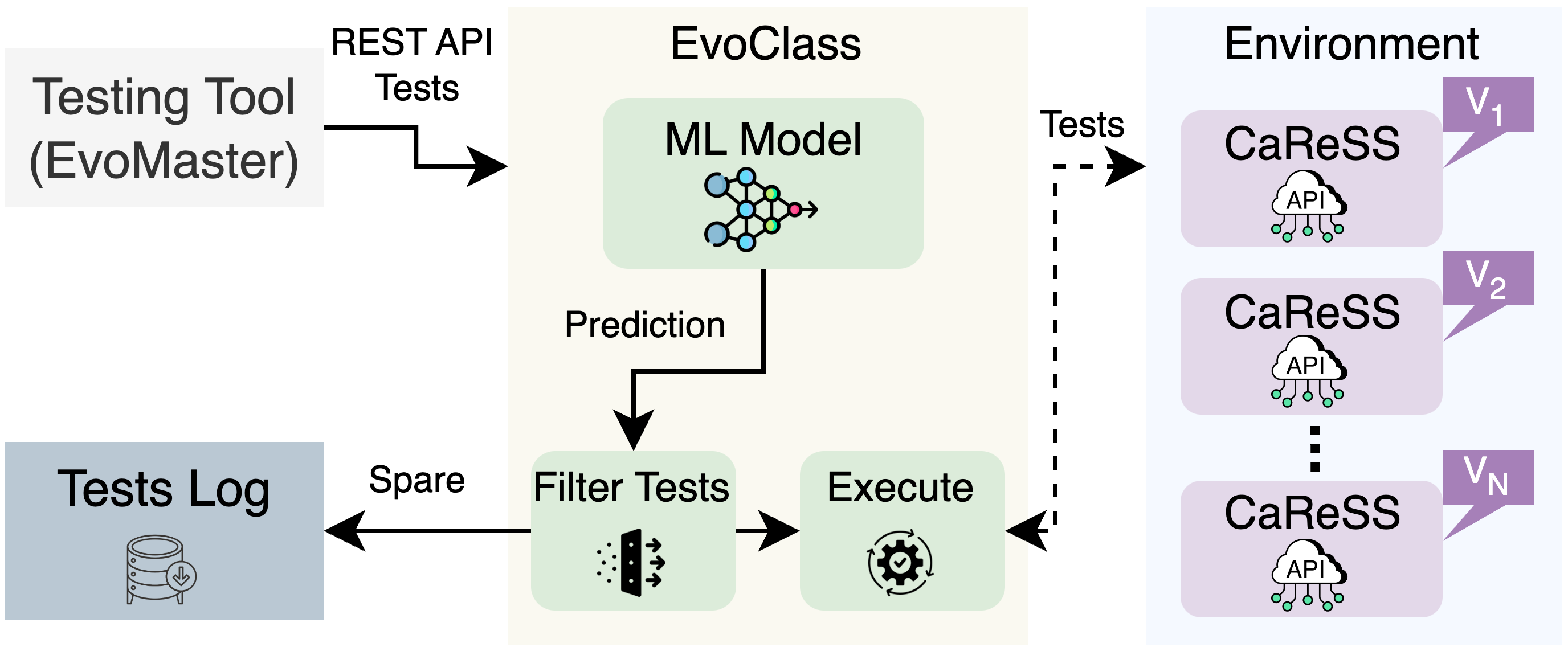}
\caption{An overview of testing CaReSS}
\label{fig:crn}
\end{figure}

CaReSS continuously evolves by adding new features, stakeholders, new/revised rules, and enhancements to the rule validation engine, resulting in multiple versions of CaReSS across different environments, such as development and testing. 
Testing each version across different environments can be both time-intensive and costly. 
To alleviate the testing cost associated with the evolving CaReSS, we proposed an ML-based approach (namely EvoClass) that predicts and filters whether a test case should be executed on the SUT~\cite{isaku2023cost}. 
As shown in Figure~\ref{fig:crn}, a testing tool such as EvoMaster~\cite{arcuri2018evomaster} generates an initial set of REST API tests. 
EvoClass uses a trained ML model to classify tests as likely to lead to success or failure. 
It discards tests with a high likelihood of producing invalid outputs, possibly due to incorrect input values. 
The refined set of tests is then executed on evolving versions of CaReSS in a particular environment.
In this work, we utilize the CaReSS rule engine dataset as one of our case studies to assess the potential of QELM in the presence of quantum noise.

\section{Experiment Design}\label{sec:experimentDesign}
\begin{figure}[!tb]
\centering
\includegraphics[width=\columnwidth]{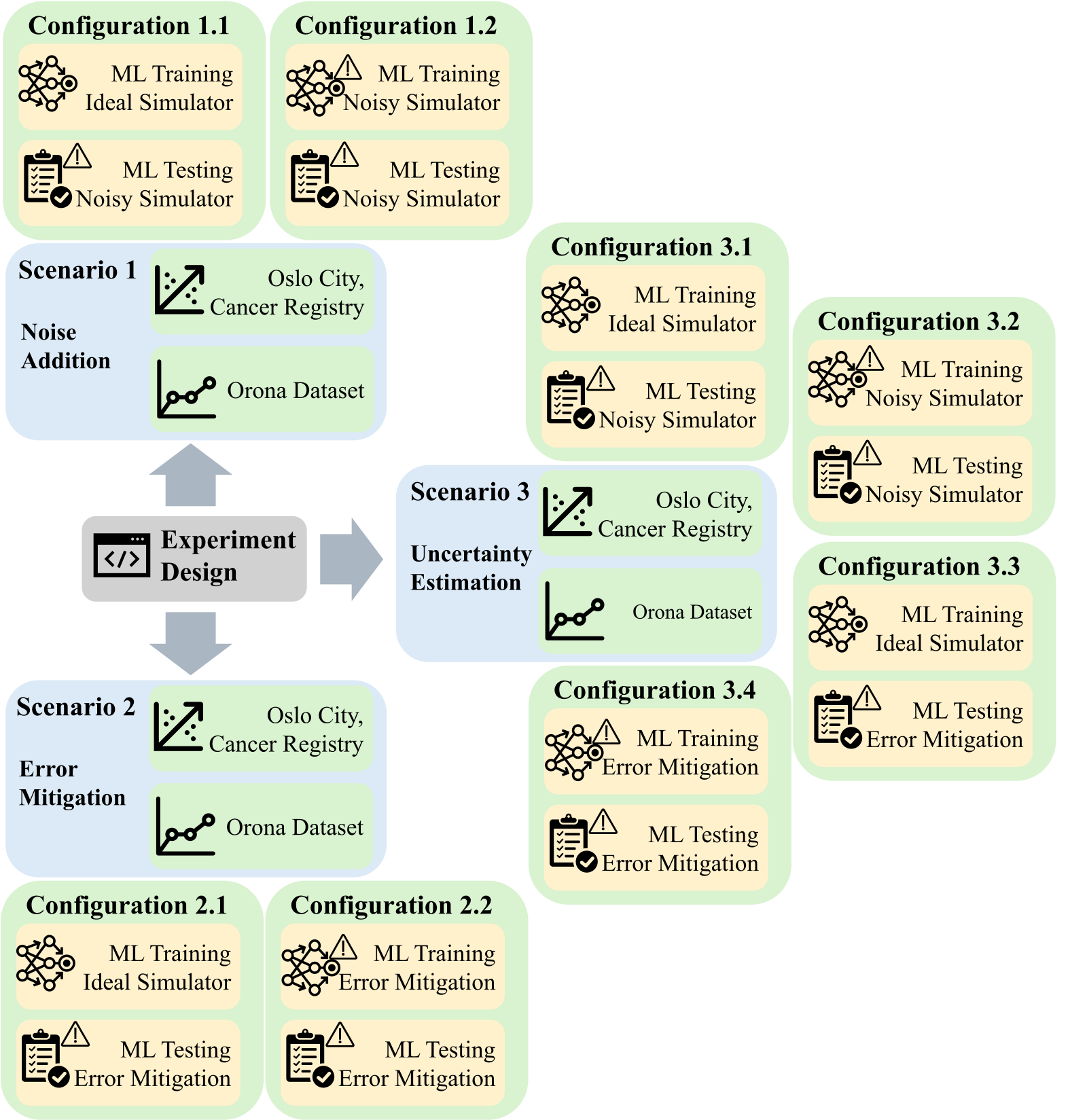}
\caption{Overview of the three scenarios evaluated to assess the performance and resistance of QELMs to quantum noise.}
\label{fig:expdesign}
\end{figure}

To evaluate the performance, resistance, and uncertainty of QELMs to quantum noise, we conducted experiments focusing on two ML tasks, i.e., regression and classification, in the context of the three industrial and real-world classical software testing approaches described in Section~\ref{sec:applicationContext}. For each task, we evaluated three distinct scenarios, each reflecting a use case relevant to the practical application of QELM. Figure~\ref{fig:expdesign} shows the overview of the three scenarios evaluated in the experiment design.

\subsection{Scenario 1 -- Noise Addition}
We evaluate the QELM's resistance against noise by introducing noise at different stages of a typical ML process. In Configuration 1.1, noise is added only during the ML testing phase, while the ML training phase is conducted on an ideal simulator. This approach evaluates situations where training can be performed on an ideal simulator, but the computational demands for time and memory are substantial. Such situations arise when the number of qubits (or features in the training data) exceeds the capacity for efficient classical simulation. In these cases, training on an ideal simulator is still feasible, albeit with high computational costs, since training is a one-time expense. However, the ML testing or deployment phase may not be conducted under ideal conditions due to high computational demands and the need for quick predictions.

In Configuration 1.2, we introduce noise in both the ML training and ML testing phases, evaluating situations where ideal simulation is not an option because the number of qubits exceeds the capacity of current simulators. Here, both training and testing must be performed on real quantum computers, providing insight into QELMs' performance in noisy environments.

\subsection{Scenario 2 -- Integration with Error Mitigation}
We explore integrating QELM with quantum error mitigation techniques since noise remains a major challenge when deploying ML models on current quantum computers. Various error mitigation strategies have been proposed to reduce the impact of this noise~\cite{mitiq}. This scenario evaluates whether combining QELMs with these techniques can enhance model accuracy and resistance to noise during both the ML training and ML testing phases. Following the same structure as in Scenario 1, we introduce quantum noise and error mitigation methods in two configurations: first, only during the ML testing phase (configuration 2.1), and then during both the ML training and ML testing phases (configuration 2.2). This approach allows us to assess whether integrating error mitigation techniques can improve the performance of QELMs for real-world applications in the current era of noisy QC.

\subsection{Scenario 3 -- Uncertainty Quantification}
While noise augmentation and error mitigation address the direct effects of quantum noise on model accuracy, an equally important consideration is the reliability of QELM predictions under these conditions. In this scenario, we incorporate Uncertainty Quantification (UQ) as a complementary evaluation layer across both noisy and error-mitigated configurations. The objective is not only to measure predictive accuracy but also to assess how confidently the QELM can make predictions when subject to noise or when error mitigation is applied. By quantifying predictive uncertainty, we can identify situations where the model's outputs are unstable or less trustworthy, even when overall accuracy appears satisfactory. This is especially important in real-world applications, where the confidence associated with a prediction can be as critical as the prediction itself.

UQ is applied to all configurations considered in Scenarios 1 and 2, resulting in four evaluation configurations:
\begin{inparaenum}[(i)]
\item ideal ML training with noisy ML testing (configuration 3.1),
\item noisy ML training and noisy ML testing (configuration 3.2),
\item ideal ML training with error-mitigated testing (configuration 3.3), and
\item error-mitigated ML training and error-mitigated ML testing (configuration 3.4).
\end{inparaenum}
This design enables us to systematically capture how noise, with or without mitigation, affects not only accuracy but also the reliability of predictions. By integrating UQ into these scenarios, we obtain a more comprehensive view of QELM performance---one that bridges accuracy, noise resilience, and confidence---thereby offering a clearer perspective on their viability in the noisy quantum computing era.

\subsection{Research Questions}
Based on the above three scenarios, we evaluate the accuracy, noise resistance, and confidence of QELMs by answering the following research questions (RQs), each considering the training and testing phase in the ML pipeline.
\begin{itemize}
\item[\textbf{\hspace{0.4em}RQ1}] How resistant is QELM to quantum noise?
\item[\textbf{\hspace{0.4em}RQ2}] How effective are current error mitigation methods for QELMs?
\item[\textbf{\hspace{0.4em}RQ3}] How does quantum noise relate to the uncertainty in QELMs?
\end{itemize}

\subsection{Dataset Characteristics}
\noindent\textbf{Orona.} 
The dataset related to the elevator dispatching algorithm used in the experiment contains 12 features, such as the number of upward and downward calls to different floors, the number of calls in the past 5 minutes, and the average travel distance. The same dataset was used in previous studies~\cite{gartziandia2022machine,quell}. The key task of the ML model employed was to predict the average waiting time for a specified period in the future. 

\noindent\textbf{Karie.}
The Karie dataset used for the experiment contains 18 features, including early access to medication, brightness settings, language preference, alarm configurations (e.g., melody, repetitions, and volume), and network connectivity. 
Note that the dataset with the same features was used to create ML-based DTs in previous work~\cite{sartaj2024medet}.
The primary function of the ML model was to predict the responses (expressed as HTTP status codes) to API requests received from an IoT application to support automated testing.

\noindent\textbf{CaReSS.}
The CaReSS dataset contains 57 features related to the patient, cancer case, and cancer message. 
These include key aspects such as patient medical records (e.g., radiation, chemotherapy, hormone treatment), cancer type, tumor behavior, stage, autopsy type, and cancer message validity. The same dataset and features were used in previous work to train a machine learning model for predicting potentially successful or unsuccessful tests~\cite{isaku2023cost}. 
The objective was to execute a minimized set of tests, thereby reducing overall testing costs.

\subsection{Benchmarks}
\noindent\textbf{QELM Models.} As described in Section~\ref{sec:qelm}, QELM models consist of three primary components: an encoder circuit, which transforms classical data into quantum states; a reservoir circuit, which is responsible for feature extraction and transformation within the quantum system; and a linear ML model, which is trained on the quantum features extracted by the reservoir circuit. Several types of encoders and reservoirs have been proposed~\cite{qelmapplication}, and the optimal combination of encoder, reservoir, and linear model depends on the specific task. Given the numerous possible combinations of encoder-reservoir circuits and linear models, evaluating all combinations under noisy conditions is impractical. 
We based our experiment on the encoder-reservoir combination recommended in a software engineering study~\cite{quell}, specifically the \textit{HE-Encoder} circuit for encoding, combined with \textit{Rotation} and \textit{Ising} circuits for the reservoir. For the linear model, we tested three widely used ML algorithms: Linear Regression, Logistic Regression, and Decision Tree Classifier. The performance of these combinations was evaluated on the three industrial datasets, and the results were compared against classical baselines from their respective studies to identify the most effective QELM configuration for each dataset. For all datasets, since the number of features exceeds 10, an ideal simulation becomes computationally expensive. The CaReSS dataset has 57 features; thus, an ideal simulation is impractical. Therefore, to establish a baseline for comparison under noisy conditions, we reduce the number of features and select the most important ones that produce results comparable to or better than the classical baseline. 
The key features were selected based on feature importance scores obtained from classical baseline models used in the respective studies. For the Orona case study, the most significant features were identified using insights from a prior QELM study on the Orona dataset~\cite{quell}. In the case of the Karie and CaReSS datasets, we applied the unbiased feature importance method based on the random forest ML model~\cite{featureimportance}, which evaluates a feature's importance by assessing its ability to produce effective splits. This feature reduction process yielded three key features for the Orona dataset, four for the Karie device dataset, and eight for the CaReSS dataset.

Table~\ref{tab:qelmconfig} presents the results of the best QELM configurations for ideal simulations across the three datasets.
\begin{table}[!tb]
\caption{Optimal QELM configuration for each case study under ideal simulation, compared with the corresponding classical baselines. The column QELM Configuration shows the best encoder, reservoir, and linear model separated by `-'.}
\centering
\begin{tabular}{ccccc}
\toprule
\textbf{Dataset} & \textbf{QELM Configuration} & \textbf{Metric} & \textbf{Score} & \textbf{Baseline} \\
\midrule
\textbf{Orona} & HE-Ising-LinearRegression & MSE & 11.12 & 15.4 \\
\textbf{Karie} & HE-Ising-DecisionTree & Accuracy & 1.0 & 0.98 \\
\textbf{CaReSS} & HE-Ising-LogisticRegression & Accuracy & 0.92 & 0.95 \\
\bottomrule
\end{tabular}
\label{tab:qelmconfig}
\end{table}
In all cases, the combination of the HE encoder circuit and Ising reservoir circuit performed the best. For linear models, linear regression was optimal for the Orona dataset, decision tree for the Karie dataset, and logistic regression for the CaReSS dataset. The {\it score} column indicates the performance of the best QELM configuration. The {\it baseline} column indicates the performance of the classical baseline of the dataset. For Orona, the classical baseline is an SVM model. For the Karie dataset, the classical baseline is MeDeT. For CaReSS, the classical baseline is EvoClass. By using only the most important features, the QELM model outperformed the classical baseline for the Orona and Karie datasets. For the CaReSS dataset, the QELM model achieved comparable accuracy to the classical baseline that utilized all 57 features. These QELM models, based on the ideal simulation, will serve as the baseline for comparing the results of noise evaluations in our experiments.
\\
\\
\noindent\textbf{Noise Models.} We chose noise models for the noisy simulations based on IBM's real quantum computers. IBM provides the noise models of their quantum computers, which approximate how quantum noise behaves on their actual devices. Specifically, we used the noise models from three IBM quantum computers: IBM Sherbrooke (Eagle r3 processor), IBM Torino (Heron r1 processor), and IBM Fez (Heron r2 processor). Each selected computer has a different quantum processor provided by IBM, enabling us to evaluate the resistance of QELMs against quantum noise under realistic conditions. All experiments were conducted using IBM's Qiskit framework~\cite{qiskit}, along with QunaSys's Quri-Paarts library~\cite{quri}, which offers the essential tools for QELM model creation and noise simulation.
\\
\\
\noindent\textbf{Error Mitigation Methods.}
Various error mitigation methods have been proposed to reduce the impact of quantum noise~\cite{mitiq}, generally falling into two categories: ML-based error mitigation~\cite{qlear,qoin,quietIEEESoftware2025,qraft} and non-ML error mitigation~\cite{mitiq}. However, each of these methods incurs additional computational costs. For our experiment, we selected one method from each category based on its practicality (in terms of feasible computational cost) and availability. For non-ML error mitigation, the most commonly adopted approach is Zero-Noise Extrapolation~(ZNE)~\cite{mitiq}, which is widely used by companies like IBM. For ML-based error mitigation, we chose the Q-LEAR method~\cite{qlear}, which is state-of-the-art and has been tested on real quantum computers.

The ZNE technique involves several hyperparameters that influence the effectiveness of error mitigation. These include the noise scaling factors, the gate folding scheme, and the extrapolation method~\cite{mitiq}. To determine the optimal configuration, we use the calibration module provided by the Mitiq library~\cite{mitiq}. This module takes a given noise model and a representative quantum circuit as input and outputs the ZNE settings that yield the best error mitigation performance for that scenario. In our experiments, across all three noise models, the optimal configuration consisted of scaling factors of 1.0, 2.0, 3.0, and 5.0, a global gate folding strategy, and a third-degree polynomial extrapolation.

\subsection{Metrics}~\label{metrics}
\noindent\textbf{RQ1 and RQ2 metrics.} RQ1 aims to assess the inherent robustness of QELMs to quantum noise, whereas RQ2 aims to assess the applicability of error mitigation methods in improving the robustness of QELMs to quantum noise. In both RQs, to evaluate performance across all datasets, we use the percentage change from ideal performance as the evaluation metric. The percentage change is calculated using the default performance metric for each dataset model—Mean Squared Error (MSE) for Orona, and Accuracy for both Karie and CaReSS. For the regression task in the Orona dataset, higher error values indicate worse performance, so we compute the percentage increase in MSE from the ideal. In contrast, for the classification tasks (Karie and CaReSS), we measure the percentage decrease in accuracy from the ideal performance. This approach ensures a consistent metric—percentage change from the ideal—for all three datasets, allowing for meaningful cross-dataset comparison. Additionally, we apply the Mann-Whitney U test and the Vargha-Delaney \Atwelve effect size, following recommendations in~\cite{recomended_effectsize,statistics2}, to assess the statistical significance of our results.
\\
\\
\noindent\textbf{RQ3 metrics.} RQ3 aims to assess the relationship between the uncertainty of QELMs and quantum noise. Since there are no UQ metrics specifically designed for QELM models, this study adopts widely used model-agnostic UQ metrics for both regression and classification tasks mentioned in Section~\ref{sec:uq}.

In regression tasks, two of the most common approaches for quantifying predictive uncertainty are 
\textit{prediction intervals} and \textit{scoring rules}~\cite{uqmethods,UQbook,scoringrule}. 

\textbf{Prediction intervals (PI)} provide a range within which the true target value is expected to lie with a specified probability $(1-\alpha)$. 
For a given input $x$, the prediction interval is defined as
\begin{equation}
\mathit{PI}_{1-\alpha}(x) = [\hat{y}_{L}(x), \hat{y}_{U}(x)],
\end{equation}
where $\hat{y}_{L}(x)$ and $\hat{y}_{U}(x)$ denote the lower and upper bounds, respectively.
\textit{Example:} A 95\% PI given by $[8.5, 11.2]$ for input $x$ contains the true outcome $y=10.7$, indicating reliable coverage~\cite{UQbook}.

To evaluate the quality of predictive distributions, scoring rules are widely employed, as they jointly account for accuracy and the sharpness of predictive uncertainty~\cite{scoringrule}. Commonly used scoring rules in regression include the Continuous Ranked Probability Score (CRPS), Check Score (CS), and Interval Score (IS).

\textbf{Continuous Ranked Probability Score (CRPS):} 
\begin{equation}
\text{CRPS}(F,y) = \int_{-\infty}^{\infty} \big(F(z) - \mathbbm{1}\{y \leq z\}\big)^2 dz,
\end{equation}
where $F$ is the predictive cumulative distribution function (CDF). CRPS defines the overall accuracy of a predicted probability distribution compared to the observed outcome.

\textbf{Check Score (CS):} Also known as the pinball loss, the CS is used for quantile predictions. For quantile level $\tau \in (0,1)$,
\begin{equation}
\text{CS}_{\tau}(y,\hat{q}_\tau) = 
\begin{cases} 
\tau (y-\hat{q}_\tau), & y \geq \hat{q}_\tau, \\[6pt]
(1-\tau)(\hat{q}_\tau-y), & y < \hat{q}_\tau,
\end{cases}
\end{equation}
where $\hat{q}_\tau$ is the predicted $\tau$-quantile. CS defines the accuracy of predicted quantiles at a given confidence level.

\textbf{Interval Score (IS):} For an interval $[\hat{y}_{L}, \hat{y}_{U}]$ at level $1-\alpha$,
\begin{equation}
\text{IS}_{\alpha}(y, \hat{y}_L, \hat{y}_U) = (\hat{y}_U - \hat{y}_L) 
+ \frac{2}{\alpha}(\hat{y}_L-y)\mathbbm{1}\{y<\hat{y}_L\} 
+ \frac{2}{\alpha}(y-\hat{y}_U)\mathbbm{1}\{y>\hat{y}_U\}.
\end{equation}
\textit{Example:} With $\alpha=0.05$, interval $[8,12]$, and $y=13$, the interval score is $IS=44$, penalizing both the interval width and the missed coverage~\cite{scoringrule}. IS defines the quality of prediction intervals, combining both their width (sharpness) and whether the true value lies inside.

For all regression metrics, lower values correspond to better performance, with zero representing the optimal case.

In classification tasks, reliability diagrams and scoring rules are commonly used to assess uncertainty~\cite{reliability,UQbook}.

\textbf{Reliability diagrams} visually compare predicted probabilities with observed frequencies. For each probability bin (e.g., $[0.8,0.9]$), one computes
\begin{equation}
\text{Observed frequency} = \frac{\# \text{ correct predictions in bin}}{\# \text{ samples in bin}},
\end{equation}
\begin{equation}
\text{Predicted probability} = \text{mean predicted confidence in bin}.
\end{equation}
A perfectly calibrated model yields points along the diagonal line~\cite{reliability}.\\
\textit{Example:} If in the bin $[0.8,0.9]$, the mean predicted confidence is $0.85$ but the actual observed accuracy is $0.75$, the model is overconfident.

Among proper scoring rules, the Brier score and logarithmic loss (cross-entropy) are most frequently employed~\cite{barrier}.

\textbf{Brier Score (BS):} For binary classification,
\begin{equation}
\text{BS} = \frac{1}{N}\sum_{i=1}^{N} (p_i - y_i)^2,
\end{equation}
where $p_i$ is the predicted probability of the positive class and $y_i \in \{0,1\}$.
\textit{Example:} If $p=0.8$ and $y=1$, then $BS = (0.8-1)^2 = 0.04$. BS is basically the mean squared error between predicted probabilities and actual outcomes.

\textbf{Logarithmic Loss (Log Loss):} The likelihood assigned to the true outcome by the predictive probability distribution given by the following equation
\begin{equation}
\text{LogLoss} = -\frac{1}{N}\sum_{i=1}^{N} \Big( y_i \log(p_i) + (1-y_i)\log(1-p_i) \Big).
\end{equation}
\textit{Example:} For $p=0.8, y=1$, $\text{LogLoss} \approx 0.223$. If $p=0.01, y=1$, then $\text{LogLoss} \approx 4.605$, reflecting severe penalty for overconfident misclassification.

Smaller values of the Brier score and log loss indicate better predictive calibration and lower uncertainty.

\noindent\textbf{Replication Package.}
We are unable to provide a replication package for this study due to the use of proprietary industrial data and the constraints of non-disclosure agreements. However, the framework utilized for building the QELM models and performing evaluations is open-source and publicly available.\footnote{\url{https://github.com/owenagnel/qreservoir}}

\section{Results}\label{sec:experimentalResults}
\subsection{RQ1 -- Resistance to Quantum Noise (Scenario 1)}

\begin{figure}[!tb]
\centering
\begin{subfigure}[b]{0.49\textwidth}
\centering
\includegraphics[width=\columnwidth]{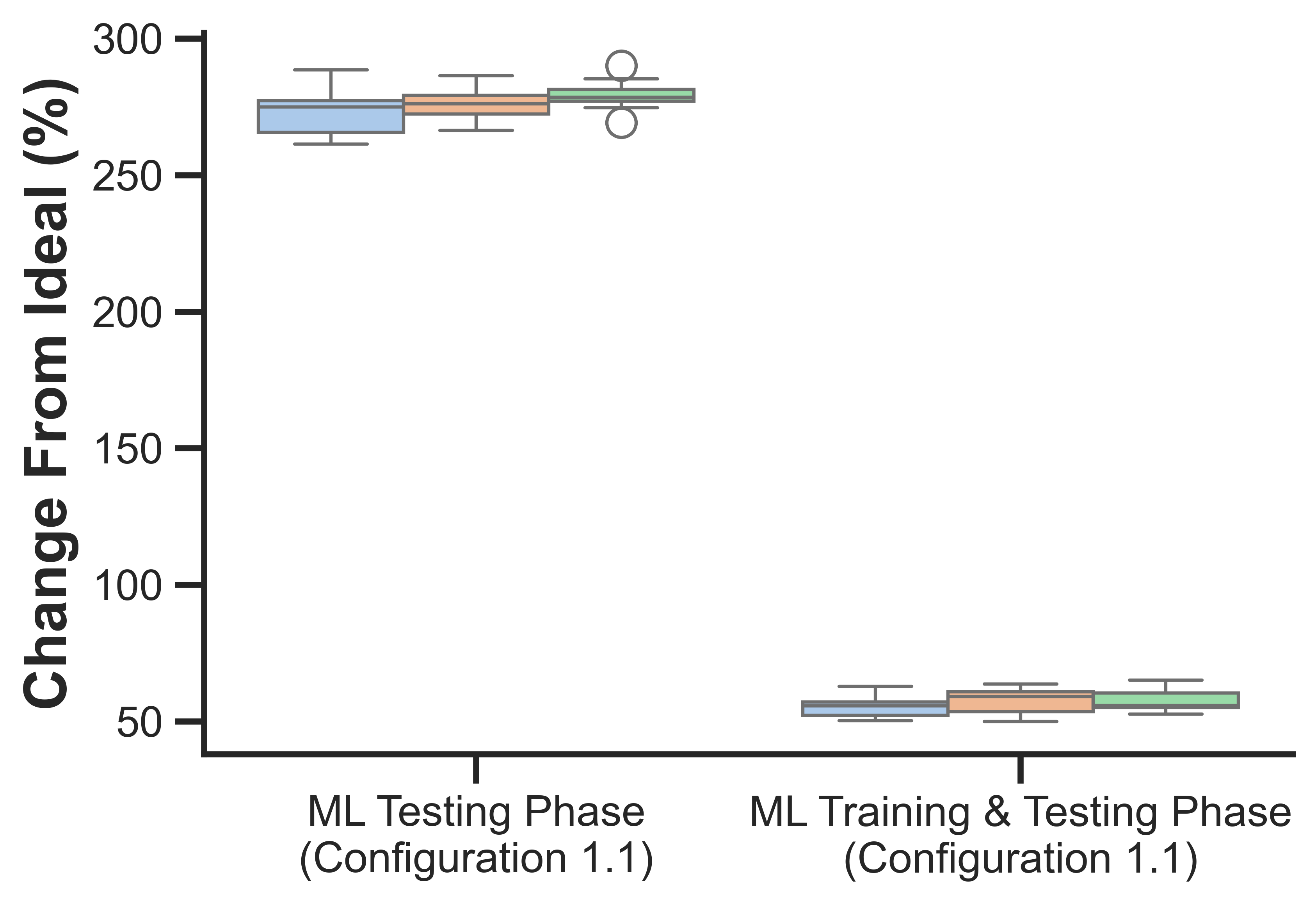}
\caption{Orona Dataset}
\label{fig:RQ1orona}
\end{subfigure}
\hfill
\begin{subfigure}[b]{0.49\textwidth}
\centering
\includegraphics[width=\columnwidth]{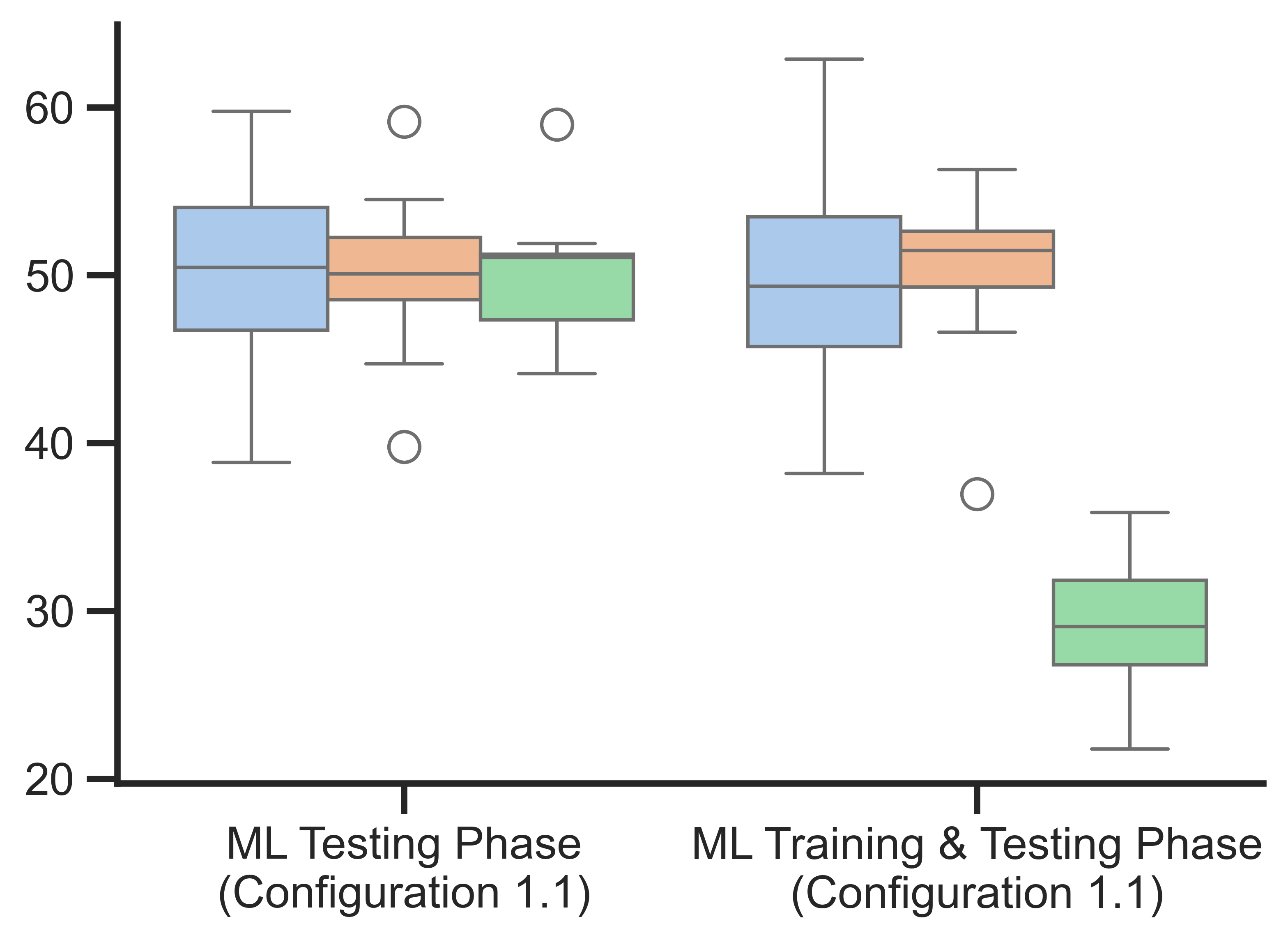}
\caption{Karie Dataset}
\label{fig:RQ1oslo}
\end{subfigure}
\hfill
\begin{subfigure}[b]{0.49\textwidth}
\centering
\includegraphics[width=\columnwidth]{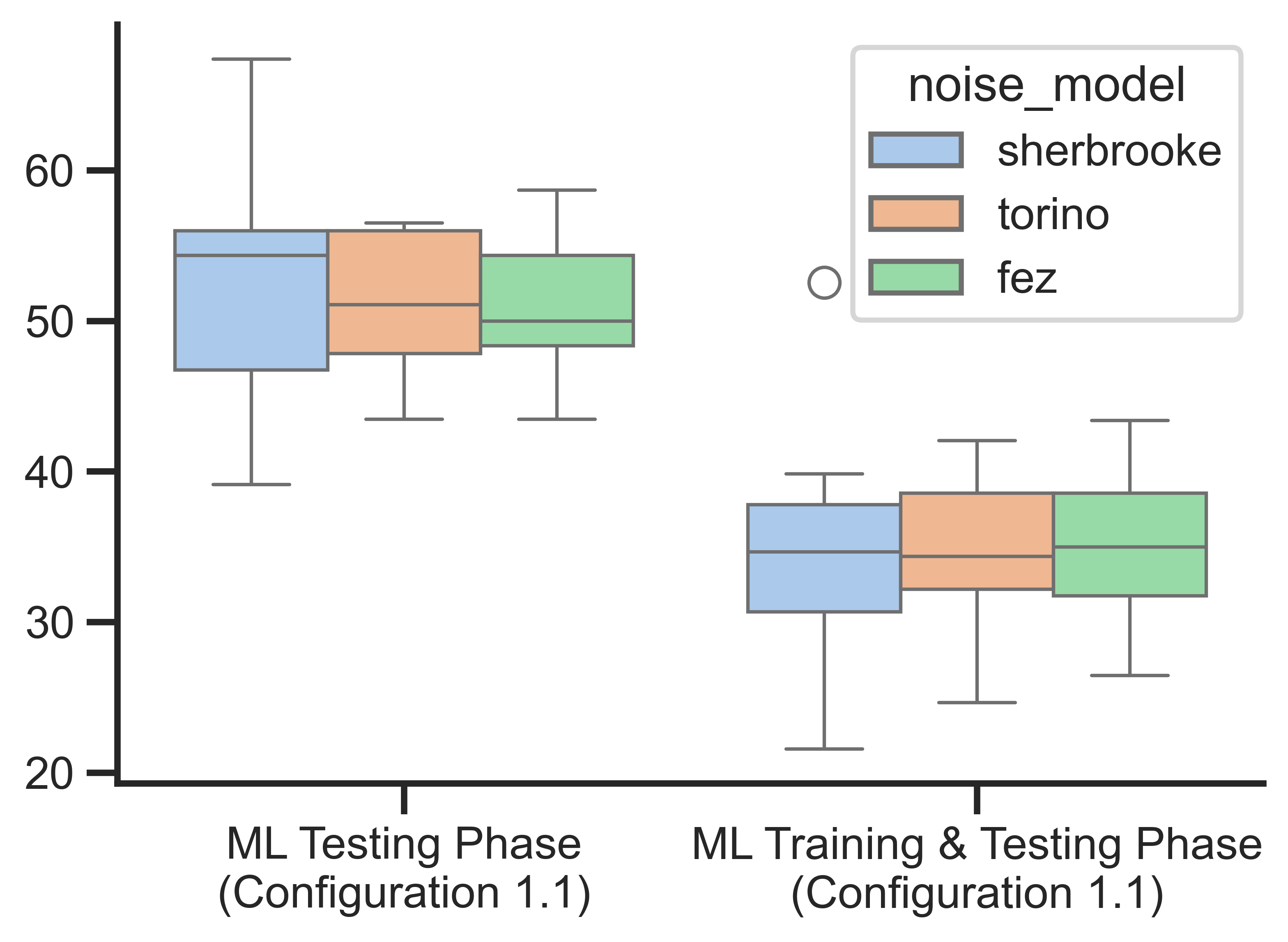}
\caption{CaReSS Dataset}
\label{fig:RQ1cancer}
\end{subfigure}
\caption{RQ1 -- Box plot showing the 10 runs for each dataset on three noise models. The $x$-axis shows the two phases of noise augmentation, and the $y$-axis shows the percentage change from the ideal values (Score) column from Table~\ref{tab:qelmconfig}.}
\label{fig:RQ1}
\end{figure}

We executed 10 times the best-performing QELM configuration for each dataset in Table~\ref{tab:qelmconfig}, across two configurations: noise introduced only during the ML testing phase (configuration 1.1), and noise present during both ML training and ML testing phases (configuration 1.2), and calculated the percentage change from the ideal performance. Additionally, we employed the Mann-Whitney statistical test and the Vargha-Delaney \Atwelve effect size, as recommended in~\cite{recomended_effectsize,statistics2}, to evaluate the statistical significance of the results across both phases. 
Figure~\ref{fig:RQ1} presents the results of 10 runs for each dataset under the three noise models.
The $x$-axis represents the two phases of noise augmentation, while the $y$-axis shows the percentage change from the ideal values for each dataset.

In Figure~\ref{fig:RQ1orona}, for the Orona dataset, the median percentage change in error exceeds 250\% for all three noise models when noise is introduced only during testing, highlighting the extremely poor performance of QELMs on the regression task. When noise is introduced in both the training and testing phases, the median percentage change decreases from over 250\% to just above 50\%, indicating that the QELM model has some ability to adapt to quantum noise when it is present during both phases. However, the deviation from the ideal performance remains significant, making it challenging to consider the model practical for real-world applications.

For the classification tasks (i.e., the Karie dataset in Figure~\ref{fig:RQ1oslo} and the CaReSS dataset in Figure~\ref{fig:RQ1cancer}), the median percentage change exceeds 50\% for all three noise models when noise is introduced only during testing. This indicates poor performance of QELMs on classification tasks, though the impact is not as severe as in the regression task. When noise is introduced during both the training and testing phases, the median percentage change decreases from over 50\% to just above 30\% for the CaReSS dataset across all three noise models. However, for the Karie dataset, no significant difference is observed for the Sherbrooke and Torino noise models, while the Fez noise model shows an improvement from above 50\% to 29\%. The fact that Sherbrooke and Torino noise models show improvements in one dataset but not in the other suggests that the ability of QELMs to learn in noisy environments for classification tasks depends on both the dataset and the noise model (the chosen quantum computer). However, even in the case of improvement, the deviation from the ideal is still too large to be practical for real-world applications.

Another observation across all three datasets is that as the number of qubits increases (3 qubits for Orona, 4 qubits for Karie, and 8 qubits for CaReSS), the variance in performance also increases. This suggests that as the number of qubits increases, the model becomes less stable, even when noise is present during both training and testing phases. This rising instability implies that QELMs struggle with noise resistance as the model size grows, and their predictions cannot be reliably trusted.

To assess the statistical significance of the results, the Mann-Whitney test was applied to each dataset for both phases across three noise models. The results for all three datasets yielded a p-value less than 0.05 with a {\it large} effect size (according to the \Atwelve classification reported in~\cite{kitchenham2017robust}), indicating that the QELM model is significantly affected by quantum noise.

Given the results, if QELMs were to be applied directly for software testing under realistic noisy quantum conditions, the instability and scalability limitations would significantly undermine trust in the testing pipeline. For Orona dataset, QELMs would fail to function as a reliable oracle. In the case of the Oslo City dataset, it would perform as an extremely poor digital twin, offering little practical value. For the Cancer Registry dataset, such performance would lead to a high incidence of false positives and false negatives, making the approach unsuitable for application in noisy conditions.

\begin{tcolorbox}[colback=blue!5!white, colframe=white, breakable]
\textbf{Answer to RQ1.} QELMs significantly struggle with quantum noise. While noise in both training and testing phases reduces its impact on model performance, the deviation from the ideal remains. The variability across datasets and noise models suggests inconsistent noise adaptability, and instability in prediction increases with the increase in qubits, indicating limited scalability and reliability of QELMs under noisy conditions. 
\end{tcolorbox}

\subsection{RQ2 -- Integration with Error Mitigation (Scenario 2)}
Similarly to RQ1, we executed 10 times the best-performing QELM configuration for each dataset in Table~\ref{tab:qelmconfig}, focusing on two configurations: error mitigation during the ML testing phase only (Configuration 2.1) and error mitigation during both ML training and ML testing phases (Configuration 2.2). We applied the same metric—percentage change from ideal—for cross-comparison between the three noise models. Our evaluation revealed that integrating error mitigation in any phase significantly reduces the variance across multiple runs, leading to lower uncertainty in the QELM models. Therefore, we report the median results from the 10 runs for better readability. Table~\ref{tab:RQ2} shows the result of integrating error mitigation methods.
\begin{table}[!tb]
\centering
\caption{RQ2 -- Result of integrating error mitigation methods in both noise augmentation phases. Column $\boldsymbol{T_N}$ shows the result of the integration of error mitigation only on the test time. Column $\boldsymbol{TT_N}$ shows the result of the integration of error mitigation on both train and test time. The row values are the median percentage change from the ideal values. Bold values show the cases where error mitigation provides significant improvement.} 
\label{tab:RQ2}
\begin{subtable}{\textwidth}
\centering
\caption{Integration with ZNE error mitigation}
\label{tab:RQ2zne}
\begin{tabular}{c|cccccc}
\toprule
\multicolumn{1}{c|}{\multirow{2}{*}{\textbf{Dataset}}} & \multicolumn{2}{c}{\textbf{Sherbrooke}} & \multicolumn{2}{c}{\textbf{Torino}} & \multicolumn{2}{c}{\textbf{Fez}} \\ \cmidrule{2-7} 
\multicolumn{1}{c|}{} & $\boldsymbol{T_N}$ & $\boldsymbol{TT_N}$ & $\boldsymbol{T_N}$ & $\boldsymbol{TT_N}$ & $\boldsymbol{T_N}$ & $\boldsymbol{TT_N}$ \\
\midrule
\textbf{Orona} & 271.8 & \textbf{9.71} & 271.8 & \textbf{12.4} & 273.1 & \textbf{1.52} \\
\textbf{Karie} & 50.0 & \textbf{2.0} & 50.0 & \textbf{1.0} & 50.0 & \textbf{0.0} \\
\textbf{CaReSS} & 56.5 & 34.78 & 56.5 & 34.78 & 56.5 & 34.78 \\
\bottomrule
\end{tabular}
\end{subtable}

\vspace{20pt}

\begin{subtable}{\textwidth}
\centering
\caption{Integration with Q-LEAR error mitigation}
\label{tab:RQ2qlear}
\begin{tabular}{c|cccccc}
\toprule
\multicolumn{1}{c|}{\multirow{2}{*}{\textbf{Dataset}}} & \multicolumn{2}{c}{\textbf{Sherbrooke}} & \multicolumn{2}{c}{\textbf{Torino}} & \multicolumn{2}{c}{\textbf{Fez}} \\ \cmidrule{2-7} 
\multicolumn{1}{c|}{} & $\boldsymbol{T_N}$ & $\boldsymbol{TT_N}$ & $\boldsymbol{T_N}$ & $\boldsymbol{TT_N}$ & $\boldsymbol{T_N}$ & $\boldsymbol{TT_N}$ \\ 
\midrule
\textbf{Orona} & 307.3 & 18.7 & 301.4 & 22.3 & 310.0 & 40.2 \\
\textbf{Karie} & 50.0 & \textbf{3.0} & 50.0 & \textbf{3.0} & 34.0 & \textbf{3.0} \\
\textbf{CaReSS} & 50.0 & \textbf{4.3} & 56.0 & \textbf{4.3} & 1.0 & \textbf{0.0} \\
\bottomrule
\end{tabular}
\end{subtable}
\end{table}
The column labeled $T_N$ displays the outcomes when error mitigation is applied only during the test phase, while column $\mathit{TT}_N$ shows the results when error mitigation is applied during both the training and testing phases. The values in the rows represent the median percentage change from the ideal values. Table~\ref{tab:RQ2zne} illustrates the results of integrating ZNE error mitigation, and Table~\ref{tab:RQ2qlear} presents the outcomes for the Q-LEAR error mitigation method.

For both ZNE and Q-LEAR error mitigation methods, the only noticeable improvement over the RQ1 results occurs when error mitigation is applied during both the training and testing phases of the QELM models. The ZNE error mitigation method improved performance under all three noise models for the Orona and Karie datasets, with median percentage changes of 9.71\%, 12.4\%, and 1.52\% for Orona. For Karie, the median percentage changes were of 2.0\%, 1.0\%, and 0.0\%, respectively; however, no improvement was observed for the CaReSS dataset. This indicates that for a number of qubits larger than 5, ZNE struggles to mitigate the noise error. For a small number of qubits, such as 3 qubits (Orona dataset), an ideal simulation is effective, and considering quantum noise is unnecessary. However, for datasets with 8 qubits (CaReSS datasets), ZNE does not improve performance, rendering it impractical for real-world applications.

The Q-LEAR error mitigation method shows significant improvement in classification datasets (Karie, CaReSS) across all noise models. For the Karie dataset, Q-LEAR shows a median percentage change of 3\% for all noise models. For the CaReSS dataset, Q-LEAR shows a median percentage change of 4.3\% for Sherbrooke and Torino noise models and 0\% for the Fez noise model. However, for the regression task, Q-LEAR underperforms compared to the ZNE method. This suggests that, unlike ZNE, Q-LEAR is less dependent on qubit size and the quantum computer used, but is more influenced by the type of problem being solved. One reason for Q-LEAR's better performance in classification tasks is due to the nature of the task itself. In classification tasks, machine learning models focus on identifying patterns that correspond to specific classes. As long as distinguishable patterns exist, ML models can learn effectively, even if the specific feature values vary. In contrast, regression tasks rely heavily on the accuracy of the feature values. For ML-based error mitigation methods like Q-LEAR, if the noise-reducing model preserves the patterns necessary for classification, the QELM model can still perform well. This is reflected in Q-LEAR's weaker performance in regression tasks, where it struggles. This suggests that Q-LEAR may alter the feature values when mitigating noise, but it retains the overall patterns for each class, benefiting classification tasks but impairing regression performance, where precise feature values are crucial.

The effectiveness of ZNE and Q-LEAR largely relies on the specific context in which they are used. ZNE's performance is constrained by the size of the qubits, whereas Q-LEAR is affected by the characteristics of the task at hand. Neither method is universally applicable to all QELM applications, highlighting the importance of developing error mitigation strategies tailored to benefit QELM models.

In the context of software testing, integrating QELMs with error mitigation during both the training and testing phases may offer a more suitable option; however, the benefits are highly dependent on the specific testing task. For example, in the Orona dataset, where the intended role of the model is to serve as an oracle, even the minimum median percentage difference of 1.52\% observed under the Fez noise model would translate into a higher rate of incorrect oracle predictions. Such deviations would ultimately compromise the reliability of the testing pipeline.

\begin{tcolorbox}[colback=blue!5!white, colframe=white, breakable]
\textbf{Answer to RQ2.} Integrating error mitigation methods enhances the noise resistance of QELMs, but their effectiveness is context-dependent. Non-ML-based methods like ZNE are constrained by qubit size, whereas ML-based methods like Q-LEAR excel in classification tasks but struggle with regression. This underscores the necessity for tailored error mitigation strategies to optimize the performance of QELM models for real-world applications.
\end{tcolorbox}

\subsection{RQ3 -- Uncertainty Quantification of QELMs}
Building on the insights from RQ1 and RQ2, we found that QELMs exhibit improved performance and robustness to quantum noise only when the noise is consistently present in both the training and testing phases, and error mitigation is applied throughout. Therefore, in RQ3, we restrict our uncertainty quantification analysis to Configuration 3.2 and Configuration 3.4 in scenario 3 from the experiment design (see Figure~\ref{fig:expdesign}). Configuration 3.2 involves applying quantum noise during both the ML training and testing phases, while Configuration 3.4 utilizes error mitigation in both phases.

\subsubsection{Regression Case (Orona Dataset)} In the context of regression tasks, calculating prediction intervals and scoring rule metrics requires a distribution of predictions. A common approach to obtain a prediction distribution is through bootstrap sampling or ensemble modeling~\cite{UQbook,uqmethods}, both of which introduce variability that helps estimate predictive uncertainty. In this study, we employ both methods to analyze and compare their effect on the uncertainty estimates of QELMs. Specifically, we use 100 bootstrap samples, where only the linear output layer is retrained while keeping the QELM reservoir weights fixed. In contrast, the ensemble approach involves training 30 independently initialized QELM models, allowing both the reservoir weights and the linear layer to vary. The motivation behind using both techniques is to examine whether changing the reservoir weights influences the level of uncertainty. Bootstrap reflects variability only in the linear model, whereas the ensemble captures variability in the full QELM architecture.

\begin{figure}[!tb]
\centering
\includegraphics[width=0.78\textwidth]{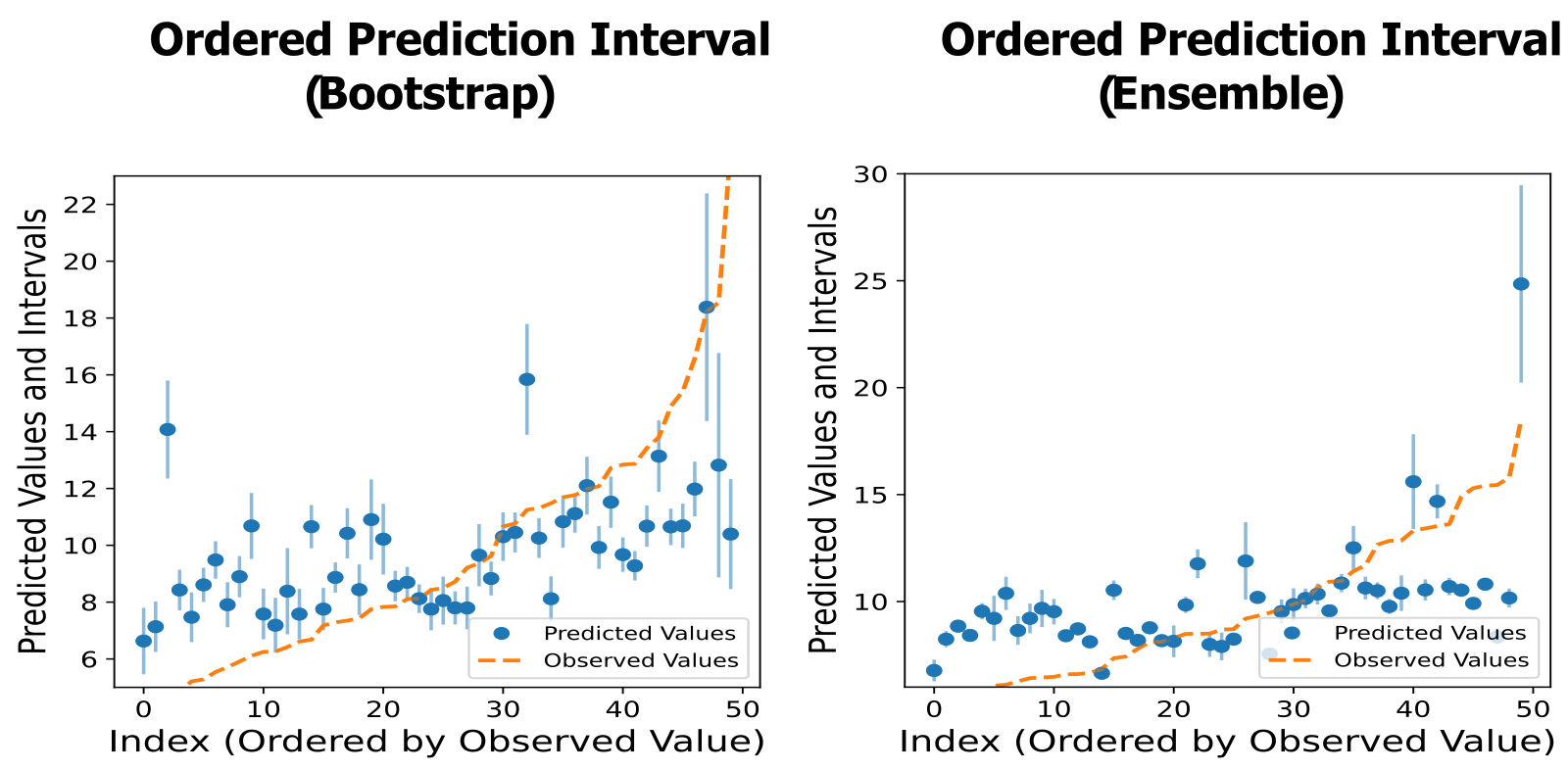}
\caption{RQ3 -- Prediction intervals for the ideal QELM model (no quantum noise) on the Orona dataset, using both bootstrap and ensemble methods. The true observed test set values are sorted and shown as an orange dashed line, while the predicted mean values are represented by solid blue dots. The 95\% prediction intervals are plotted around the predicted means.}
\label{fig:orona_idealUQ}
\end{figure}

Figure~\ref{fig:orona_idealUQ} presents the prediction intervals of the ideal QELM model—i.e., the version without quantum noise—for the Orona dataset, using both bootstrap and ensemble methods.
These intervals serve as a reference baseline, allowing us to compare how uncertainty levels change—either increasing or decreasing—when quantum noise is introduced and when error mitigation techniques are applied.

\begin{figure}[!tb]
\centering
\includegraphics[width=\textwidth]{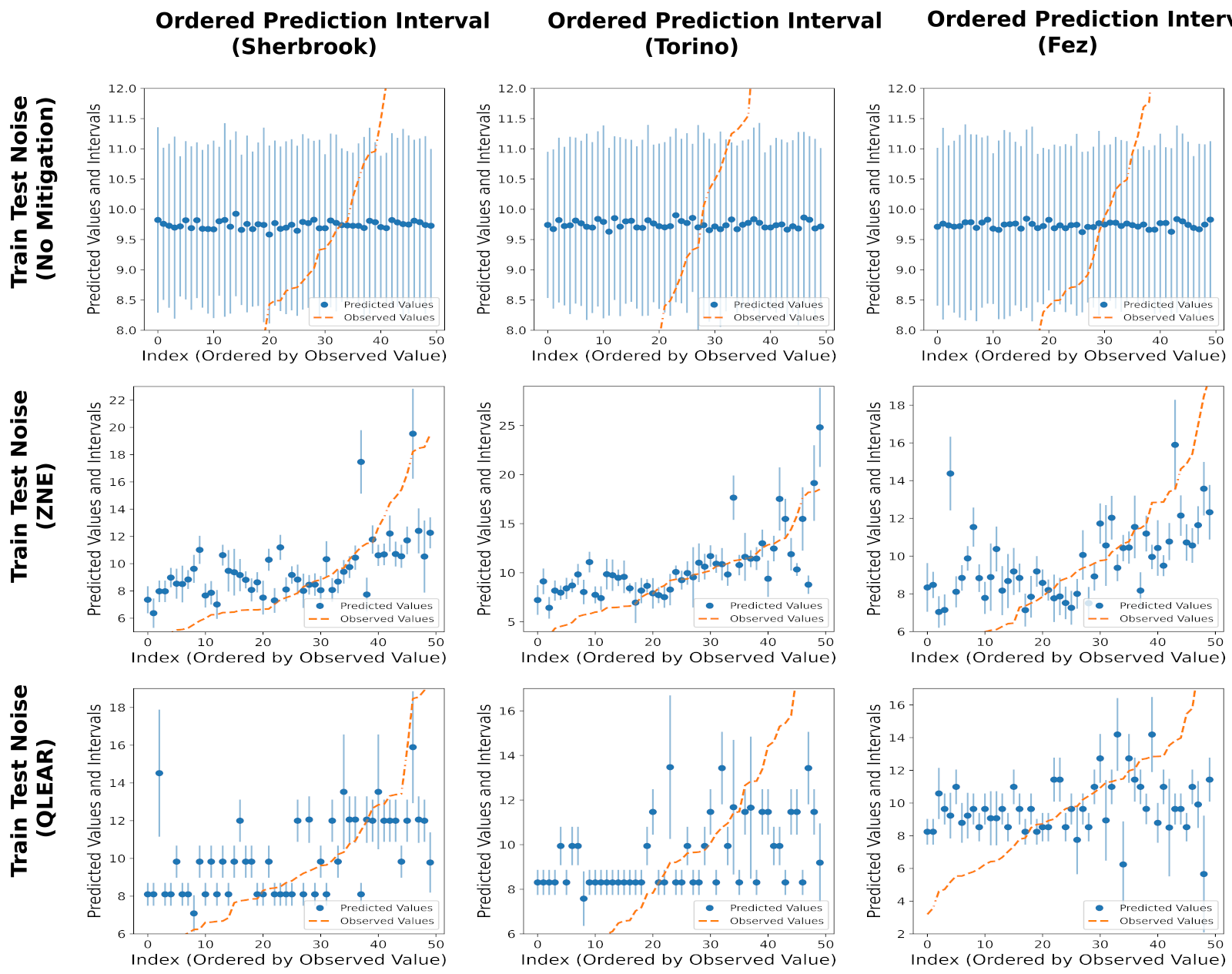}
\caption{RQ3 -- Bootstrap sampling Prediction intervals on the Orona dataset for IBM noise models, ZNE, and QLEAR method. The true observed test set values are sorted and shown as an orange dashed line, while the predicted mean values are represented by solid blue dots. The 95\% prediction intervals are plotted around the predicted means.}
\label{fig:orona_boostrapUQ}
\end{figure}

\begin{figure}[!tb]
\centering
\includegraphics[width=\textwidth]{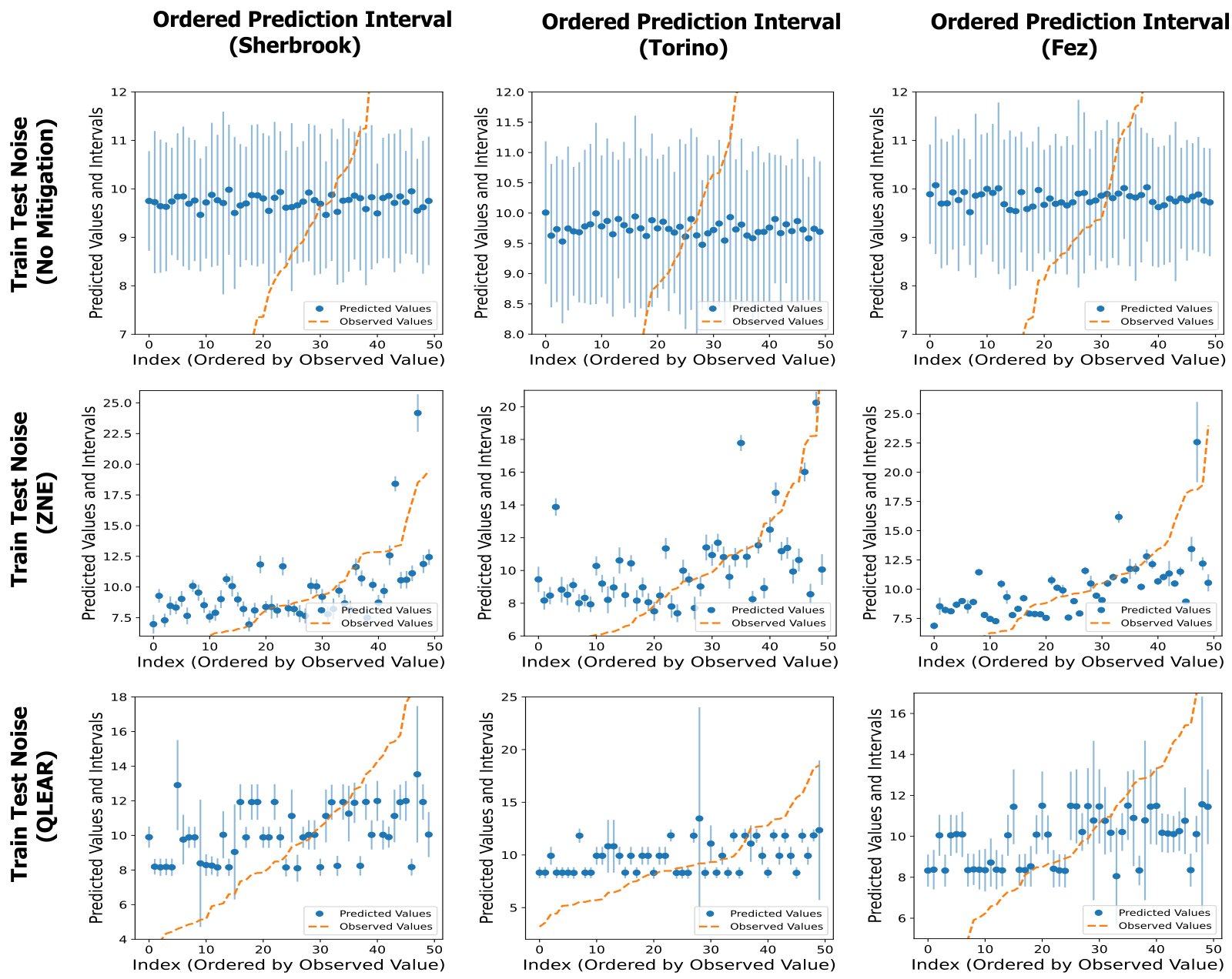}
\caption{RQ3 -- Ensemble sampling Prediction intervals on the Orona dataset for IBM noise models, ZNE, and QLEAR method. The true observed test set values are sorted and shown as an orange dashed line, while the predicted mean values are represented by solid blue dots. The 95\% prediction intervals are plotted around the predicted means.}
\label{fig:orona_ensembleUQ}
\end{figure}

Figures~\ref{fig:orona_boostrapUQ} and~\ref{fig:orona_ensembleUQ} illustrate the prediction intervals for the Orona dataset under three configurations:
\begin{inparaenum}[(1)]
\item quantum noise applied during both training and testing phases (Train-Test Noise),
\item Zero Noise Extrapolation applied in both phases (Train-Test Noise ZNE), and
\item QLEAR applied in both phases (Train-Test Noise QLEAR), using both bootstrap and ensemble methods across all three IBM noise models.
\end{inparaenum}

In the Train-Test Noise setting without mitigation, both Figures~\ref{fig:orona_boostrapUQ} and \ref{fig:orona_ensembleUQ} reveal that quantum noise causes a compression effect on the QELM outputs—most predicted means cluster within a narrow range (9–10)—and results in significantly wider 95\% prediction intervals. This suggests that predictions made under unmitigated noise are highly uncertain and unreliable, regardless of their correctness. With ZNE, however, prediction intervals across all noise models closely resemble those of the ideal (noise-free) case shown in Figure~\ref{fig:orona_idealUQ}. Notably, the ensemble method yields tighter confidence bounds than the bootstrap, indicating that QELMs combined with ZNE and ensemble strategies can achieve both accurate and more trustworthy predictions under noise. In contrast, the QLEAR configuration shows some reduction in uncertainty compared to the no-mitigation case, but the improvement is limited. Moreover, its prediction intervals tend to be wider than those with ZNE, suggesting that QLEAR may be less effective for regression tasks. This observation is consistent with findings from RQ2, where ZNE proved more effective as an error mitigation strategy for regression.

Table~\ref{tab:RQ3orona} presents the results of the scoring rule metrics discussed in Section~\ref{metrics}.
\begin{table}[!tb]
\centering
\caption{RQ3 -- Scoring Rule metrics for Orona dataset for both bootstrap and ensemble distributions. Column $\mathit{TT}_N$, $TT_Z$, $\mathit{TT}_Q$ shows the value for Train-Test noise no mitigation, Train-Test noise ZNE, and Train-Test noise QLEAR, respectively. The closet value to the ideal for each metric is shown in bold.}
\label{tab:RQ3orona}
\begin{subtable}{\textwidth}
\centering
\caption{Distribution generated with Bootstrap Sampling}
\label{tab:RQ3oronaboot}
\resizebox{\textwidth}{!}{%
\begin{tabular}{c|c|ccccccccc}
\toprule
\multicolumn{1}{c|}{\multirow{2}{*}{\textbf{Metric}}} & \multicolumn{1}{c|}{\multirow{2}{*}{\textbf{Ideal}}} & \multicolumn{3}{c}{\textbf{Sherbrooke}} & \multicolumn{3}{c}{\textbf{Torino}} & \multicolumn{3}{c}{\textbf{Fez}} \\ \cmidrule{3-11} 
\multicolumn{1}{c|}{} & \multicolumn{1}{c|}{} & $\boldsymbol{TT_N}$ & $\boldsymbol{TT_Z}$ & $\boldsymbol{TT_Q}$ & $\boldsymbol{TT_N}$ & $\boldsymbol{TT_Z}$ & $\boldsymbol{TT_Q}$ & $\boldsymbol{TT_N}$ & $\boldsymbol{TT_Z}$ & $\boldsymbol{TT_Q}$ \\
\midrule
\textbf{CRPS} & 2.29 & 3.00 & \textbf{2.39} & 2.61 & 3.00 & \textbf{2.35} & 2.68 & 3.00 & \textbf{2.29} & 2.84 \\
\textbf{Check Score} & 1.15 & 1.50 & \textbf{1.20} & 1.31 & 1.50 & \textbf{1.18} & 1.34 & 1.51 & \textbf{1.15} & 1.42 \\
\textbf{Interval Score} & 19.8 & 25.5 & \textbf{20.5} & 23.3 & 25.4 & \textbf{19.7} & 23.9 & 25.6 & \textbf{19.6} & 24.5 \\
\bottomrule
\end{tabular}%
}
\end{subtable}

\vspace{20pt}

\begin{subtable}{\textwidth}
\centering
\caption{Distribution generated with Ensemble Sampling}
\label{tab:RQ3oronaensemble}
\resizebox{\textwidth}{!}{%
\begin{tabular}{c|c|ccccccccc}
\toprule
\multicolumn{1}{c|}{\multirow{2}{*}{\textbf{Metric}}} & \multicolumn{1}{c|}{\multirow{2}{*}{\textbf{Ideal}}} & \multicolumn{3}{c}{\textbf{Sherbrooke}} & \multicolumn{3}{c}{\textbf{Torino}} & \multicolumn{3}{c}{\textbf{Fez}} \\ \cmidrule{3-11} 
\multicolumn{1}{c|}{} & \multicolumn{1}{c|}{} & $\boldsymbol{TT_N}$ & $\boldsymbol{TT_Z}$ & $\boldsymbol{TT_Q}$ & $\boldsymbol{TT_N}$ & $\boldsymbol{TT_Z}$ & $\boldsymbol{TT_Q}$ & $\boldsymbol{TT_N}$ & $\boldsymbol{TT_Z}$ & $\boldsymbol{TT_Q}$ \\
\midrule
\textbf{CRPS} & 2.44 & 3.02 & \textbf{2.53} & 2.61 & 3.03 & \textbf{2.53} & 2.69 & 3.00 & \textbf{2.46} & 2.64 \\
\textbf{Check Score} & 1.22 & 1.51 & \textbf{1.27} & 1.30 & 1.52 & \textbf{1.27} & 1.35 & 1.50 & \textbf{1.23} & 1.32 \\
\textbf{Interval Score} & 23.1 & 25.9 & 23.6 & \textbf{23.4} & 26.2 & \textbf{23.7} & 24.1 & 25.7 & 23.8 & \textbf{22.0} \\
\bottomrule
\end{tabular}%
}
\end{subtable}
\end{table}
Specifically, Table~\ref{tab:RQ3oronaboot} reports results using the bootstrap method, while Table~\ref{tab:RQ3oronaensemble} shows results for the ensemble approach. In both cases, the Train-Test Noise configuration without any mitigation ($\mathit{TT}_N$) leads to a clear increase in all three metrics—CRPS, Check Score, and Interval Score—across all three IBM noise models, indicating higher uncertainty and less reliable predictions compared to the ideal (noise-free) baseline.

For the ZNE configuration ($TT_Z$), the bootstrap results in Table~\ref{tab:RQ3oronaboot} show that all three metrics are much closer to the ideal values across all noise models, demonstrating that ZNE effectively reduces uncertainty while improving predictive reliability. The QLEAR configuration ($\mathit{TT}_Q$) also shows improvement over $\mathit{TT}_N$ in all metrics for all noise models in the bootstrap case, but its performance does not match that of ZNE, suggesting a more limited impact.

In the ensemble results (Table~\ref{tab:RQ3oronaensemble}), ZNE again performs best, achieving near-ideal values in two out of the three metrics. QLEAR also shows improvement, but is closer to the ideal in only one metric. However, the difference between ZNE and QLEAR is relatively small, indicating that both methods enhance uncertainty estimates, though ZNE appears to be the more effective option for regression tasks.

\subsubsection{Classification Case (Oslo City, Cancer Registry Dataset) }
\begin{figure}[!tb]
\includegraphics[width=\textwidth]{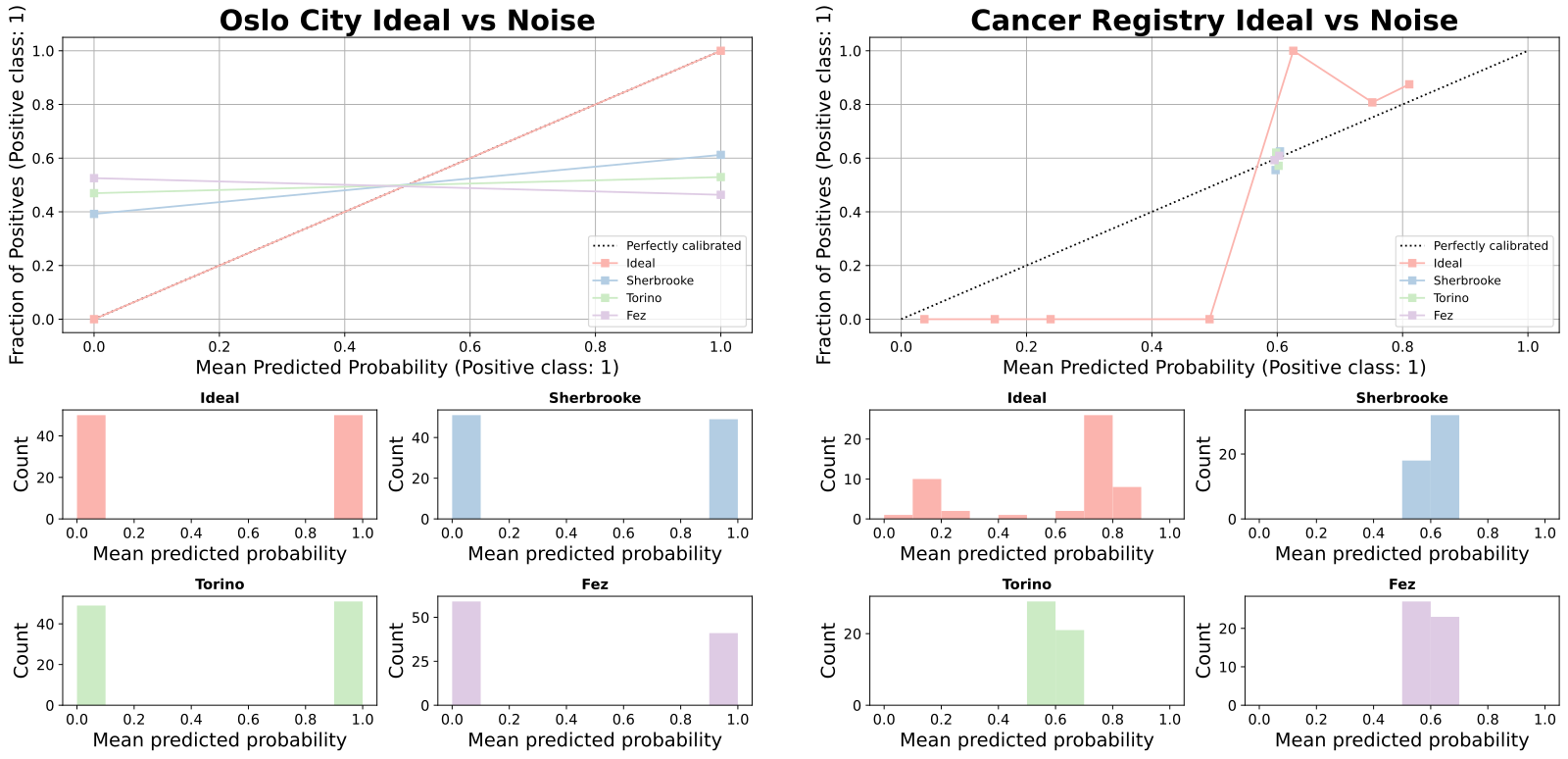}
\caption{RQ3 -- Reliability diagram for Oslo City and Cancer Registry in comparison to the noise models. The line chart displays the reliability diagram, while the bottom histogram illustrates the distribution of predicted probabilities.}
\label{fig:oslocityUQ}
\end{figure}

To assess uncertainty in the classification datasets (Oslo City and Cancer Registry), we applied reliability diagrams, barrier score, and log loss as evaluation metrics. Figure~\ref{fig:oslocityUQ} presents the reliability plots for both datasets under all three noise models, alongside the ideal reference.
The line plot illustrates the reliability diagram, while the histogram below depicts the distribution of the model's predictions. On the line plot, the x-axis represents the mean predicted probability, reflecting the model's confidence. For instance, a point at 0.8 corresponds to predictions where the model estimated 80\% confidence. The y-axis indicates the fraction of positives, representing the actual outcomes—e.g., for 80\% predicted confidence, what proportion were truly positive. A perfectly calibrated model aligns with the diagonal dotted line. Models positioned above this line are overconfident, while those below are underconfident.

For the Oslo City dataset, the ideal model is perfectly calibrated, demonstrating that it reliably translates confidence into accurate predictions. Its histogram further indicates that predictions are made with extreme confidence, concentrated at 0\% and 100\%. In contrast, under quantum noise, the QELM models display a saturating behavior: across all confidence levels, their accuracy remains around only 40–50\%, making them highly unreliable. Even more concerning, the histograms for all three noise models show predictions clustered at extreme confidence levels, revealing that the models not only perform poorly but also make incorrect predictions with strong confidence. For the Cancer Registry dataset, the ideal model is not perfectly calibrated, exhibiting a mix of overconfident and underconfident predictions. However, under quantum noise, the model demonstrates saturation not only in accuracy but also in confidence. This indicates that as the qubit count increases, quantum noise rapidly erodes the learning capability of QELM models, rendering them highly unreliable.

\begin{figure}[!tb]
\includegraphics[width=\textwidth]{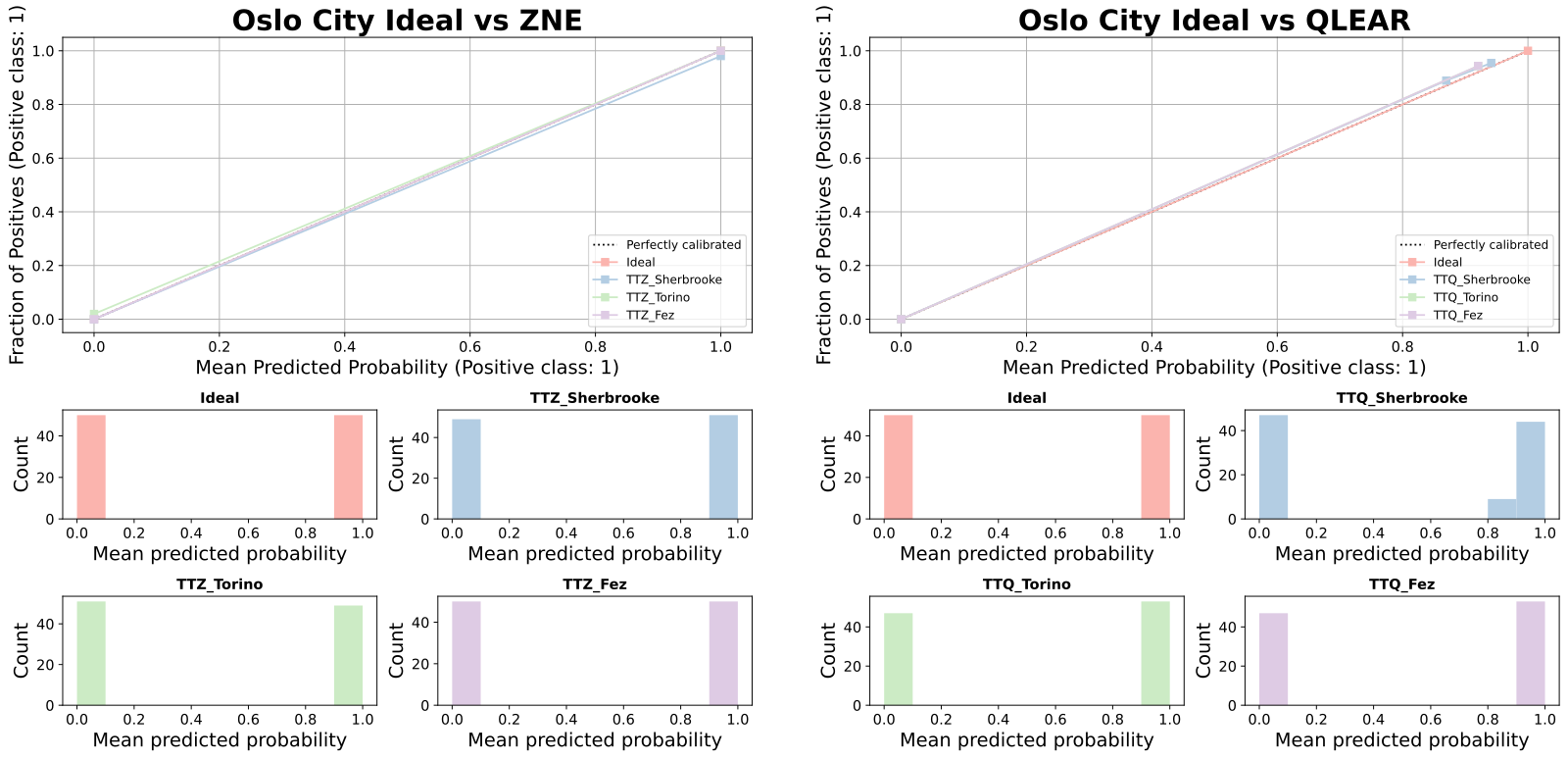}
\caption{RQ3 -- Reliability diagram for Oslo City combined with ZNE and QLEAR error mitigation. The line chart displays the reliability diagram, while the bottom histogram illustrates the distribution of predicted probabilities.}
\label{fig:oslocityerrUQ}
\end{figure}

Figure~\ref{fig:oslocityerrUQ} illustrates the integration of ZNE and QLEAR for the Oslo City dataset.
In terms of uncertainty, both methods bring the reliability diagram closer to the ideal model, indicating that they effectively reduced uncertainty. However, QLEAR introduces a slight shift in the confidence distributions across all three noise models, as seen in the histogram. While the difference is small compared to the ideal case, it may reflect the inherent uncertainty of QLEAR itself, given that it is also a machine learning–based error mitigation technique.

\begin{figure}[!tb]
\includegraphics[width=\textwidth]{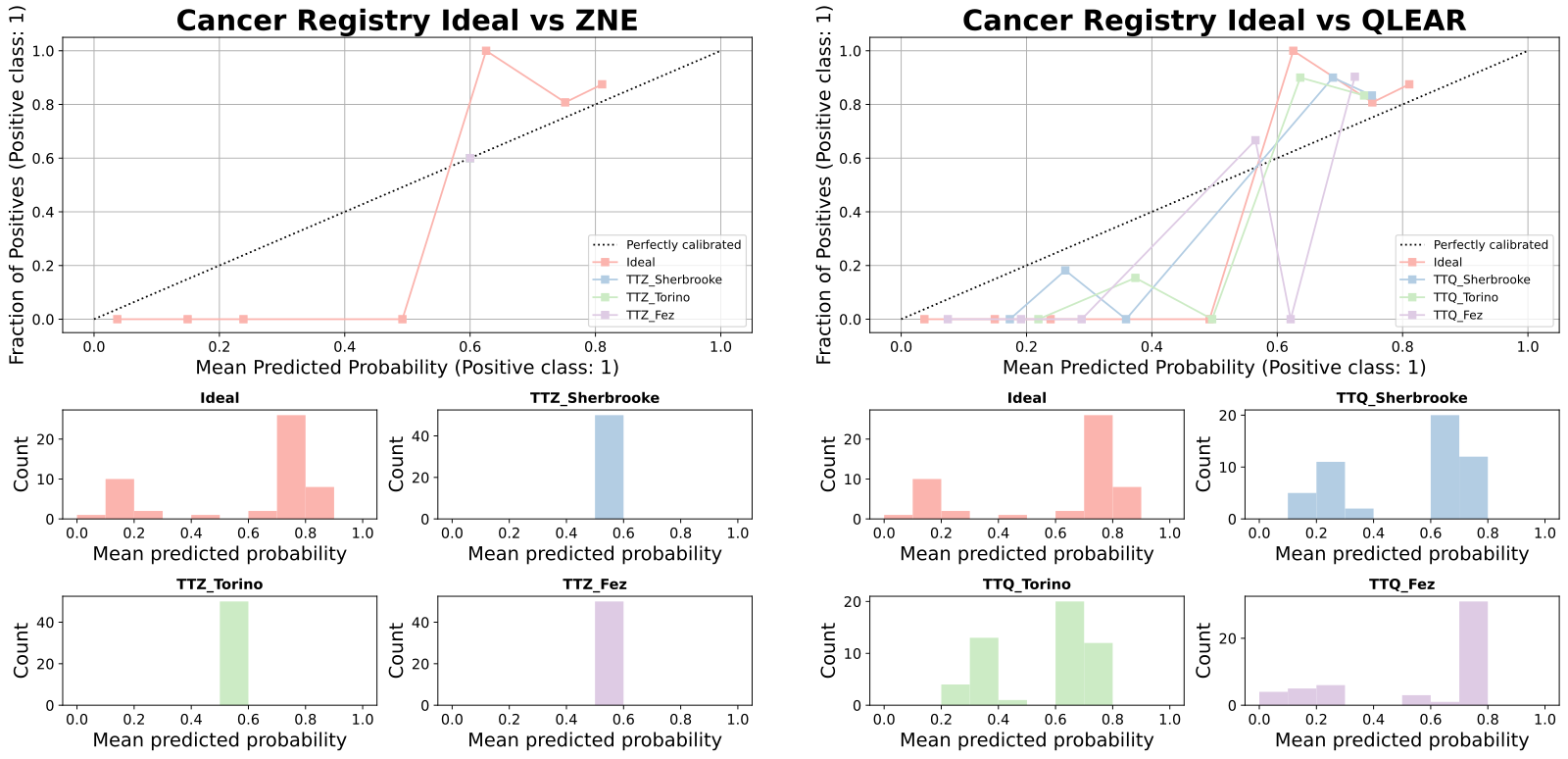}
\caption{RQ3 -- Reliability diagram for Cancer Registry combined with ZNE and QLEAR error mitigation. The line chart displays the reliability diagram, while the bottom histogram illustrates the distribution of predicted probabilities.}
\label{fig:cancererrUQ}
\end{figure}

Figure~\ref{fig:cancererrUQ} presents the integration of ZNE and QLEAR for the Cancer Registry dataset.
For ZNE, the results show no improvement in reducing uncertainty; in fact, the reliability diagram becomes even more saturated than under pure quantum noise. This suggests that as the number of qubits increases, ZNE becomes less effective for classification tasks, consistent with the findings from RQ2. In contrast, QLEAR successfully reduces uncertainty to levels comparable with the ideal model, indicating that for classification tasks, QLEAR may serve as a more effective error mitigation approach.

For the scoring rule metrics (barrier score and log loss), Table~\ref{tab:RQ3oslocancer} reports the results for both the Oslo City and Cancer Registry datasets.
\begin{table}[!tb]
\centering
\caption{RQ3 -- Scoring Rule metrics for both Oslo City and Cancer Registry datasets. Columns $\mathit{TT}_N$, $\mathit{TT}_Z$, $\mathit{TT}_Q$ show the value for Train-Test noise no mitigation, Train-Test noise ZNE, and Train-Test noise QLEAR, respectively. The closet value to the ideal for each metric is shown in bold.}
\label{tab:RQ3oslocancer}
\begin{subtable}{\textwidth}
\centering
\caption{Oslo City Dataset}
\label{tab:RQ3oslo}
\resizebox{\textwidth}{!}{%
\begin{tabular}{c|c|ccccccccc}
\toprule
\multicolumn{1}{c|}{\multirow{2}{*}{\textbf{Metric}}} & \multicolumn{1}{c|}{\multirow{2}{*}{\textbf{Ideal}}} & \multicolumn{3}{c}{\textbf{Sherbrooke}} & \multicolumn{3}{c}{\textbf{Torino}} & \multicolumn{3}{c}{\textbf{Fez}} \\ \cmidrule{3-11} 
\multicolumn{1}{c|}{} & \multicolumn{1}{c|}{} & $\boldsymbol{TT_N}$ & $\boldsymbol{TT_Z}$ & $\boldsymbol{TT_Q}$ & $\boldsymbol{TT_N}$ & $\boldsymbol{TT_Z}$ & $\boldsymbol{TT_Q}$ & $\boldsymbol{TT_N}$ & $\boldsymbol{TT_Z}$ & $\boldsymbol{TT_Q}$ \\
\midrule
\textbf{Barrier Score} & 0 & 0.39 & \textbf{0.01} & 0.02 & 0.47 & \textbf{0.01} & 0.02 & 0.53 & \textbf{0} & 0.02 \\
\textbf{Log Loss} & 0 & 14.05 & 0.36 & \textbf{0.11} & 16.94 & 0.36 & \textbf{0.11} & 19.10 & \textbf{0} & 0.11 \\ \hline
\end{tabular}%
}
\end{subtable}

\vspace{20pt}

\begin{subtable}{\textwidth}
\centering
\caption{Cancer Registry Dataset}
\label{tab:RQ3cancer}
\resizebox{\textwidth}{!}{%
\begin{tabular}{c|c|ccccccccc}
\toprule
\multicolumn{1}{c|}{\multirow{2}{*}{\textbf{Metric}}} & \multicolumn{1}{c|}{\multirow{2}{*}{\textbf{Ideal}}} & \multicolumn{3}{c}{\textbf{Sherbrooke}} & \multicolumn{3}{c}{\textbf{Torino}} & \multicolumn{3}{c}{\textbf{Fez}} \\ \cmidrule{3-11} 
\multicolumn{1}{c|}{} & \multicolumn{1}{c|}{} & $\boldsymbol{TT_N}$ & $\boldsymbol{TT_Z}$ & $\boldsymbol{TT_Q}$ & $\boldsymbol{TT_N}$ & $\boldsymbol{TT_Z}$ & $\boldsymbol{TT_Q}$ & $\boldsymbol{TT_N}$ & $\boldsymbol{TT_Z}$ & $\boldsymbol{TT_Q}$ \\
\midrule
\textbf{Barrier Score} & 0.11 & 0.23 & 0.24 & \textbf{0.12} & 0.24 & 0.24 & \textbf{0.15} & 0.23 & 0.24 & \textbf{0.10} \\
\textbf{Log Loss} & 0.38 & 0.67 & 0.67 & \textbf{0.43} & 0.67 & 0.67 & \textbf{0.49} & 0.67 & 0.67 & \textbf{0.38} \\ \hline
\end{tabular}%
}
\end{subtable}
\end{table}
The column $\mathit{TT}_N$ shows the baseline results under noise without error mitigation, while $\mathit{TT}_Z$ and $\mathit{TT}_Q$ represent the outcomes with ZNE and QLEAR integration, respectively. Values closest to the ideal baseline are highlighted in bold. For Oslo City (Table~\ref{tab:RQ3oslo}), ZNE performs slightly better than QLEAR in terms of barrier score, whereas QLEAR achieves lower log loss. This suggests that both methods are effective in reducing uncertainty in QELM for this dataset. For the Cancer Registry (Table~\ref{tab:RQ3cancer}), however, ZNE shows no improvement over the noise model, while QLEAR consistently reduces uncertainty and brings the results closer to the ideal baseline. These findings align with the reliability diagrams, reinforcing that ZNE becomes less effective as the number of qubits increases.

Overall, the results indicate that although error mitigation can bring model accuracy closer to the ideal, a higher-accuracy yet highly uncertain model would still have limited practical value in software testing. Its instability would undermine reliability despite strong average performance. In the Orona oracle task, uncertainty implies that identical system behaviors could alternately be flagged as correct or faulty, eroding trust in the oracle's judgments and necessitating repeated human validation. For the Oslo City digital twin, inconsistency across runs would result in an unpredictable digital twin: in some cases, the model may closely mirror real behavior, while in others, it may diverge drastically, rendering it unsuitable as a reliable testbed. In the Cancer Registry rule-filtering task, instability would appear as fluctuating false positives and false negatives---either overwhelming testers with spurious defect reports or, more critically, failing to detect genuine issues. Thus, while improved accuracy suggests potential, the absence of stability across these tasks ultimately renders QELMs ineffective for real-world software testing pipelines under the effect of quantum noise.

\begin{tcolorbox}[colback=blue!5!white,colframe=white,breakable]
\textbf{Answer to RQ3.} Quantum noise amplifies uncertainty in QELM models, which rises with qubit count before plateauing at a high level—a saturation effect where accumulated noise overwhelms useful information and drives outputs towards a fixed value. Error mitigation can partially curb this effect, with QLEAR proving more effective for classification and ZNE for regression and smaller qubit settings. Yet, no method consistently controls uncertainty across all tasks or scales.
\end{tcolorbox}

\section{Threats to validity}\label{sec:threats}
This section discusses the threats to validity across four established categories~\cite{threats}: construct validity, internal validity, conclusion validity, and external validity.
\subsection{Construct Validity}
A potential threat lies in the metrics used to assess the noise resistance of QELMs. We addressed this by using the percentage change from the ideal QELM model score as the comparison metric, where the ideal score is calculated based on the same metrics used in the original case studies for classical baselines. This allows for a fair comparison of classical baseline ML models with QELM models. Another potential threat arises from the fact that, to the best of our knowledge, there are currently no uncertainty quantification (UQ) methods specifically developed for Quantum Extreme Learning Machines (QELMs). This absence makes it difficult to evaluate predictive uncertainty in a manner that fully reflects the unique characteristics of quantum models, such as their inherently probabilistic nature and representation within Hilbert space, which differs fundamentally from the Gaussian assumptions often underlying classical machine learning models. To address this limitation, we employed model-agnostic UQ metrics that are widely used in both regression and classification tasks. While these metrics allow for fair comparison across modeling approaches, they are not tailored to QELMs and may therefore fail to capture aspects of uncertainty that are intrinsic to quantum-based models.

\subsection{Internal Validity}
A potential threat arises from the QELM configuration used across all three datasets. To mitigate this, we selected the best-performing configuration of the encoder-reservoir-linear model for each dataset under ideal conditions (no quantum noise) based on the study~\cite{quell} and a preliminary experiment. We evaluated this optimal setup under noisy conditions. Real-world applications aim to maximize performance, so evaluating the best configuration helps to understand how well the QELM model performs under noise in practical scenarios. 

\subsection{Conclusion Validity}
In the case of QELM models, the introduction of quantum noise adds inherent randomness, which can lead to variability in outcomes. To mitigate this issue and reduce the impact of random bias, we conducted ten repeated runs for research questions one and two in order to draw more reliable conclusions. Another threat arises in research question three, where the uncertainty metrics require probability distributions estimated across multiple runs. To address this, we employed two widely used model-agnostic distribution generation methods: bootstrap sampling and ensemble modeling. In our study, we utilized 100 bootstrap samples and 30 ensemble models. These values were selected based on the practical feasibility of executing noisy quantum simulations on classical hardware within a reasonable time frame. While increasing the number of bootstrap samples or ensemble models might yield slightly different results, we believe the overall conclusions would remain consistent.

\subsection{External Validity}
One challenge is the selection of case studies. We chose realistic industrial case studies where classical ML models are already in use, demonstrating that QELMs can potentially replace them for improved performance and accuracy. Another threat concerns the quantum noise model employed in our experiments. Quantum systems are subject to various sources of noise, and noise models can be constructed in multiple ways to approximate these effects~\cite{noise_benchmark1}. This introduces the risk that the chosen noise model may not fully capture all relevant error processes, thereby limiting the validity of our results. To mitigate this threat, we utilized IBM's noise models, which are specifically designed to reflect the behavior of their real quantum devices. These models are maintained and updated regularly by IBM, providing a closer approximation to the actual noise characteristics of current quantum hardware.

\section{Discussion and Lesson Learnt}\label{sec:discussion}
\subsection{QELM Applications in Software Engineering}
QELMs offer potential advantages in performance and accuracy for tasks where classical machine learning models are typically used for classification and regression. Consequently, this applies to applying classical machine learning techniques for classical software engineering tasks---a widely studied topic in recent years~\cite{mlforse, Kotti2023}. Software engineering tasks, such as requirements engineering, design and modeling, implementation, testing, and project management, all benefit from classical machine learning models. Specific tasks like requirements tracing, user story detection, automated software modeling, code smell detection, test case generation and optimization, project cost estimation, and performance prediction~\cite{mlforse} are prime examples. In all these tasks, QELMs could potentially replace classical machine learning models, delivering improved performance and efficiency. Nonetheless, further investigation is needed.

However, in the current era of noisy quantum computing, QELMs are only practical for small-scale problems. For tasks requiring fewer qubits, such as those involving 1-8 features, QELMs can be simulated on ideal quantum simulators, enabling noise-free execution and better results than classical models. As task complexity increases and the number of qubits grows, ideal simulations become infeasible. Our findings indicate that QELMs are not sufficiently resistant to quantum noise, preventing them from reaching their full potential on real quantum computers. This limitation hinders the scalability of QELMs, making them less viable for large-scale industrial applications in software engineering. To tackle these challenges, a promising approach is to explore methods for breaking complex problems into smaller subproblems, enabling ideal computations that can be managed more effectively with fewer qubits. Numerous techniques from classical computing, such as graph partitioning~\cite{graph} and hierarchical decomposition~\cite{heiraricy}, can be investigated and adapted for use. Additionally, developing enhanced encoding methods could allow for greater information compression into fewer qubits. This direction aligns with ongoing research in QML~\cite{dataenc}, and our future work will also focus on these strategies in the context of software testing.

\subsection{Practical limitations}
For QELM models to be practically applicable to larger problem sizes, using real quantum computers is essential, making error mitigation techniques crucial. However, our study shows that the QELM performance with error mitigation is highly context-dependent, varying with the application and the quantum computer used. Additionally, error mitigation methods introduce significant computational overhead. Both ML-based and non-ML-based methods require multiple quantum circuit executions to calculate their respective features to mitigate noise, which dramatically increases computational costs as problem sizes grow. For instance, the fastest method in our study, ZNE, requires at least three repeated executions per circuit, tripling the computational cost for predicting a single data point. This time cost can negate the advantages of using QELMs, limiting their practical use. In specialized areas, such as medical or safety-critical systems, where either quantum dynamics play an important role or small improvements can yield significant benefits, QELMs with different error mitigation methods to find a working combination on real quantum computers may be worth exploring. However, for more general applications, deploying QELMs with current quantum technology will require customized strategies and task-specific error mitigation techniques.

\subsection{Uncertainty Quantification}
In our exploration of Uncertainty Quantification (UQ) for QELMs, we found that model-specific methods developed for classical machine learning are not directly transferable, due to fundamental differences between the quantum and classical domains. Model-specific UQ methods are advantageous because they exploit the mathematical structure and inductive biases of the underlying model, enabling them to generate more accurate, efficient, and interpretable uncertainty estimates compared to model-agnostic approaches~\cite{UQbook}. Unlike generic resampling or ensemble-based techniques, which treat the model as a black box and often yield conservative bounds, model-specific methods can disentangle epistemic uncertainty (arising from limited data, insufficient parameters, or under-sampling) from aleatoric uncertainty (inherent noise in the data), while remaining consistent with the model's training objective. This distinction has proven especially valuable in the classical counterparts of QELMs, such as extreme learning machines and reservoir computing~\cite{guerra2023probabilistic, domingo2024quantifying}.

For QELMs, however, generic UQ methods, such as those relying on Gaussian approximations, struggle to capture true predictive uncertainty. Since QELMs leverage the expressivity of quantum states and operate in high-dimensional Hilbert spaces, their uncertainty structure is tightly linked to the quantum feature mapping and the random feature construction. Model-agnostic techniques disregard this internal structure and risk misrepresenting uncertainty: either by producing overly loose bounds or by becoming computationally prohibitive, as excessive sampling would be required to approximate the distribution accurately. Thus, QELMs demand model-specific approaches capable of capturing the unique interplay between quantum variability, model expressivity, and data noise, ensuring that uncertainty estimates are both meaningful and practically useful.

One promising direction is the integration of quantile regression into QELM predictions, analogous to recent work in classical reservoir computing~\cite{guerra2023probabilistic}. In future research, we aim to further investigate model-driven UQ strategies tailored specifically for QELMs, with the goal of developing methods that exploit quantum structure to deliver more reliable and interpretable uncertainty estimates.

\section{Related Work}\label{sec:related}
QML has emerged as a promising field that harnesses quantum computing to enhance classical ML models, with applications across numerous domains. Researchers have systematically investigated QML's potential to outperform classical methods in various tasks. A recent survey of 94 papers on QML applications~\cite{slrqml} identified the most commonly used algorithms as quantum neural networks (QNN), quantum kernel models (QKM), variational quantum eigensolver (VQE), quantum approximate optimization algorithm (QAOA), and quantum annealing. These algorithms show promise in fields such as image processing, natural language processing, software engineering~\cite{quantOptForSESurvey2025,WangRoadMap2025,ZhaoQBSE2025}, and physics simulations~\cite{slrqml}. Specifically, in software engineering, quantum annealing~\cite{bootqa,bqtmizerIEEESoftware2025,trovato2024reformulating} and QAOA~\cite{xinyiqaoa} have been applied for test case minimization, and QELM is utilized for regression testing in software systems~\cite{quell}. Despite the broader exploration of QML, limited studies have focused on its performance in the presence of quantum noise. Previous research~\cite{qelm,qelm1,qelm3} has examined the adaptability of QML to quantum noise and compared it to classical ML algorithms. One study~\cite{qelm3} found that VQE-based QML models could perform classification tasks under noise in the IoT domain. Other studies~\cite{qelm,qelm1} concluded that QELMs demonstrate greater resistance to quantum noise than algorithms like VQE and QAOA, which suffer from issues like barren plateaus that hinder their learning capacity.

Most existing research focuses on ideal quantum simulations, with limited studies exploring the effects of noise. The studies that examine the quantum noise effect~\cite{qelm, qelm3,qelm1} do not consider practical industrial applications but instead provide insight into the general learning capacity of QML models under noisy conditions. In contrast, in our study, we conduct experiments to evaluate the real-world applicability of QELMs for software engineering industrial applications in the current era of noisy quantum computers.

In the context of Uncertainty Quantification (UQ), to the best of our knowledge, no prior work has specifically addressed UQ for QELMs. By contrast, UQ in classical machine learning has been extensively studied, with recent surveys mapping both methodological advances and domain-specific applications. For instance, \cite{he2025} proposed a taxonomy that distinguishes between data (aleatoric) and model (epistemic) uncertainty, providing guidance on method selection for tasks such as active learning and robustness in deep neural networks. Similarly, \cite{gawlikowski2023survey} reviewed a wide range of techniques—including Bayesian neural networks, ensembles, and calibration strategies highlighting their applications across medicine, robotics, and earth sciences. In more specialized contexts, \cite{siddique2022survey} demonstrated the importance of UQ in space physics, while \cite{lopez2025} surveyed UQ applications in healthcare pipelines, emphasizing its role in ensuring reliability and trust in clinical machine learning.

Since no QELM-specific UQ methods are currently available, our study adopts model-agnostic approaches, which can be applied irrespective of the underlying architecture.

\section{Conclusion}\label{sec:conclusion}
Quantum Extreme Learning Machines (QELMs) are a novel quantum machine learning approach that enhances information processing through quantum-mechanical principles. Recent research highlights their potential across various fields, including software engineering. This paper evaluated the practical application of QELMs under realistic quantum noise conditions across three industrial case studies in classical software testing in practice. Our results show that QELMs perform well in ideal simulations; however, they are significantly affected by quantum noise, limiting their practical utility on current noisy quantum computers. Although augmenting quantum noise during training and testing improves performance, the impact of noise remains too substantial for effective real-world use. Due to quantum noise, the predictive uncertainty in QELM also rises. Our results show that as the number of qubits increases, the uncertainty increases and eventually saturates to a fixed output, driving QELM models toward unreliability. Error mitigation techniques can enhance noise resistance and reduce uncertainty, but their effectiveness varies by context. Non-ML-based methods depend on factors like qubit size and the specific quantum hardware, whereas ML-based methods perform well in classification but struggle with regression tasks. This highlights the need for QELM-specific error mitigation strategies to enhance their applicability in noisy quantum environments.


\section*{Declarations}

\subsection*{Acknowledgments}
We acknowledge the use of IBM Quantum services for this work. The views expressed are those of the authors and do not reflect the official policy or position of IBM or the IBM Quantum team.

\subsection*{Funding}
Qu-Test project (Project \#299827) funded by the Research Council of Norway supports this work. Shaukat Ali is also supported by Oslo Metropolitan University's Quantum Hub and Simula's internal strategic project on quantum software engineering. This work is also partially supported by the WTT4Oslo project (No. 309175) and the AIT4CR project (No. 309642), funded by the Research Council of Norway. Paolo Arcaini is supported by the ASPIRE grant No. JPMJAP2301, JST. Aitor Arrieta is part of the Software and Systems Engineering research group of Mondragon Unibertsitatea (IT1519-22), supported by the Department of Education, Universities and Research of the Basque Country.

\subsection*{Data Availability Statement}
We are unable to provide a replication package for this study due to the use of proprietary industrial data and the constraints of non-disclosure agreements. However, the framework utilized for building the QELM models and performing evaluations is open-source and publicly available.\footnote{\url{https://github.com/owenagnel/qreservoir}}


\begin{thebibliography}{10}
\providecommand{\url}[1]{#1}
\csname url@samestyle\endcsname
\providecommand{\newblock}{\relax}
\providecommand{\bibinfo}[2]{#2}
\providecommand{\BIBentrySTDinterwordspacing}{\spaceskip=0pt\relax}
\providecommand{\BIBentryALTinterwordstretchfactor}{4}
\providecommand{\BIBentryALTinterwordspacing}{\spaceskip=\fontdimen2\font plus
\BIBentryALTinterwordstretchfactor\fontdimen3\font minus \fontdimen4\font\relax}
\providecommand{\BIBforeignlanguage}[2]{{%
\expandafter\ifx\csname l@#1\endcsname\relax
\typeout{** WARNING: IEEEtran.bst: No hyphenation pattern has been}%
\typeout{** loaded for the language `#1'. Using the pattern for}%
\typeout{** the default language instead.}%
\else
\language=\csname l@#1\endcsname
\fi
#2}}
\providecommand{\BIBdecl}{\relax}
\BIBdecl

\bibitem{mlforse}
S.~Wang, L.~Huang, A.~Gao, J.~Ge, T.~Zhang, H.~Feng, I.~Satyarth, M.~Li, H.~Zhang, and V.~Ng, ``Machine/deep learning for software engineering: A systematic literature review,'' \emph{IEEE Transactions on Software Engineering}, vol.~49, no.~3, pp. 1188--1231, 2023.

\bibitem{Kotti2023}
Z.~Kotti, R.~Galanopoulou, and D.~Spinellis, ``Machine learning for software engineering: A tertiary study,'' \emph{ACM Comput. Surv.}, vol.~55, no.~12, Mar. 2023.

\bibitem{qml}
J.~Biamonte, P.~Wittek, N.~Pancotti, P.~Rebentrost, N.~Wiebe, and S.~Lloyd, ``Quantum machine learning,'' \emph{Nature}, vol. 549, no. 7671, pp. 195--202, 2017.

\bibitem{xinyiqaoa}
X.~Wang, S.~Ali, T.~Yue, and P.~Arcaini, ``Quantum approximate optimization algorithm for test case optimization,'' \emph{IEEE Transactions on Software Engineering}, vol.~50, no.~12, pp. 3249--3264, 2024.

\bibitem{quell}
X.~Wang, S.~Ali, A.~Arrieta, P.~Arcaini, and M.~Arratibel, ``Application of quantum extreme learning machines for {QoS} prediction of elevators' software in an industrial context,'' in \emph{Companion Proceedings of the 32nd ACM International Conference on the Foundations of Software Engineering}, ser. FSE 2024.\hskip 1em plus 0.5em minus 0.4em\relax New York, NY, USA: Association for Computing Machinery, 2024, pp. 399--410.

\bibitem{qc}
J.~Preskill, ``Quantum computing in the {NISQ} era and beyond,'' \emph{Quantum}, vol.~2, p.~79, 2018.

\bibitem{qelmapplication}
K.~Nakajima and I.~Fischer, \emph{Reservoir computing}.\hskip 1em plus 0.5em minus 0.4em\relax Springer, 2021.

\bibitem{softwaretestingchallenge}
\BIBentryALTinterwordspacing
Z.~Khaliq, S.~U. Farooq, and D.~A. Khan, ``Artificial intelligence in software testing: Impact, problems, challenges and prospect,'' \emph{CoRR}, vol. abs/2201.05371, 2022. [Online]. Available: \url{https://arxiv.org/abs/2201.05371}
\BIBentrySTDinterwordspacing

\bibitem{binary_class}
K.~Fujii and K.~Nakajima, \emph{Quantum Reservoir Computing: A Reservoir Approach Toward Quantum Machine Learning on Near-Term Quantum Devices}.\hskip 1em plus 0.5em minus 0.4em\relax Singapore: Springer Singapore, 2021, pp. 423--450.

\bibitem{simulators_are_slow}
E.~Younis, K.~Sen, K.~Yelick, and C.~Iancu, ``{{QFAST}}: Conflating search and numerical optimization for scalable quantum circuit synthesis,'' in \emph{2021 IEEE International Conference on Quantum Computing and Engineering (QCE)}, 2021, pp. 232--243.

\bibitem{simulationlimit}
Y.~Zhou, E.~M. Stoudenmire, and X.~Waintal, ``{What Limits the Simulation of Quantum Computers?}'' \emph{Phys. Rev. X}, vol.~10, p. 041038, Nov 2020.

\bibitem{karie}
AceAge, ``{Karie Medicine Dispenser},'' \url{https://kariehealth.com/}, 2024, [Online; accessed 20-Aug-2025].

\bibitem{sartaj2023testing}
H.~Sartaj, S.~Ali, T.~Yue, and K.~Moberg, ``{Testing Real-World Healthcare IoT Application: Experiences and Lessons Learned},'' in \emph{Proceedings of the 31st ACM Joint European Software Engineering Conference and Symposium on the Foundations of Software Engineering}, ser. ESEC/FSE 2023.\hskip 1em plus 0.5em minus 0.4em\relax New York, NY, USA: Association for Computing Machinery, 2023, pp. 2044--2049.

\bibitem{sartaj2024medet}
H.~Sartaj, S.~Ali, and J.~M. Gjøby, ``{MeDeT: Medical Device Digital Twins Creation with Few-shot Meta-learning},'' \emph{ACM Transactions on Software Engineering and Methodology}, vol.~34, no.~6, pp. 1--36, 2025.

\bibitem{isaku2023cost}
E.~Isaku, H.~Sartaj, C.~Laaber, T.~Yue, S.~Ali, T.~Schwitalla, and J.~F. Nygård, ``Cost reduction on testing evolving cancer registry system,'' in \emph{2023 IEEE International Conference on Software Maintenance and Evolution (ICSME)}, 2023, pp. 508--518.

\bibitem{mitiq}
R.~LaRose, A.~Mari, S.~Kaiser, P.~J. Karalekas, A.~A. Alves, P.~Czarnik, M.~E. Mandouh, M.~H. Gordon, Y.~Hindy, A.~Robertson, P.~Thakre, M.~Wahl, D.~Samuel, R.~Mistri, M.~Tremblay, N.~Gardner, N.~T. Stemen, N.~Shammah, and W.~J. Zeng, ``Mitiq: A software package for error mitigation on noisy quantum computers,'' \emph{Quantum}, vol.~6, p. 774, Aug 2022.

\bibitem{qlear}
A.~Muqeet, S.~Ali, T.~Yue, and P.~Arcaini, ``A machine learning-based error mitigation approach for reliable software development on {IBM}'s quantum computers,'' in \emph{Companion Proceedings of the 32nd ACM International Conference on the Foundations of Software Engineering}, ser. FSE 2024.\hskip 1em plus 0.5em minus 0.4em\relax New York, NY, USA: Association for Computing Machinery, 2024, pp. 80--91.

\bibitem{qoin}
A.~Muqeet, T.~Yue, S.~Ali, and P.~Arcaini, ``Mitigating noise in quantum software testing using machine learning,'' \emph{IEEE Transactions on Software Engineering}, vol.~50, no.~11, pp. 2947--2961, 2024.

\bibitem{quietIEEESoftware2025}
A.~Muqeet, S.~Ali, and P.~Arcaini, ``{QUIET}: A tool for sampling-based quantum noise error mitigation,'' \emph{IEEE Software}, pp. 1--6, 2025.

\bibitem{dirac}
P.~A.~M. Dirac, ``A new notation for quantum mechanics,'' in \emph{Mathematical Proceedings of the Cambridge Philosophical Society}, vol.~35, no.~3.\hskip 1em plus 0.5em minus 0.4em\relax Cambridge University Press, 1939, pp. 416--418.

\bibitem{noise_benchmark1}
S.~Resch and U.~R. Karpuzcu, ``{Benchmarking Quantum Computers and the Impact of Quantum Noise},'' \emph{ACM Comput. Surv.}, vol.~54, no.~7, jul 2021.

\bibitem{decoherence_def}
R.~Alicki, ``{Decoherence and the Appearance of a Classical World in Quantum Theory},'' \emph{Journal of Physics A: Mathematical and General}, vol.~37, no.~5, p. 1948, feb 2004.

\bibitem{crosstalkgatenoise}
T.~Ayral, F.-M.~L. R{\'e}gent, Z.~Saleem, Y.~Alexeev, and M.~Suchara, ``{Quantum divide and compute: exploring the effect of different noise sources},'' \emph{SN Computer Science}, vol.~2, no.~3, p. 132, 2021.

\bibitem{simulators}
A.~Jamadagni, A.~M. L{\"{a}}uchli, and C.~Hempel, ``Benchmarking quantum computer simulation software packages,'' \emph{CoRR}, vol. abs/2401.09076, 2024.

\bibitem{elmsurvey}
J.~Wang, S.~Lu, S.-H. Wang, and Y.-D. Zhang, ``A review on extreme learning machine,'' \emph{Multimedia Tools and Applications}, vol.~81, no.~29, pp. 41\,611--41\,660, Dec 2022.

\bibitem{qelm}
L.~Innocenti, S.~Lorenzo, I.~Palmisano, A.~Ferraro, M.~Paternostro, and G.~M. Palma, ``Potential and limitations of quantum extreme learning machines,'' \emph{Communications Physics}, vol.~6, no.~1, p. 118, May 2023.

\bibitem{UQbook}
T.~Gneiting, F.~Balabdaoui, and A.~E. Raftery, ``Probabilistic forecasts, calibration and sharpness,'' \emph{Journal of the Royal Statistical Society Series B: Statistical Methodology}, vol.~69, no.~2, pp. 243--268, 03 2007.

\bibitem{uqmethods}
K.~Tran, W.~Neiswanger, J.~Yoon, Q.~Zhang, E.~Xing, and Z.~W. Ulissi, ``Methods for comparing uncertainty quantifications for material property predictions,'' \emph{Machine Learning: Science and Technology}, vol.~1, no.~2, p. 025006, 2020.

\bibitem{qmlbenifits}
C.~Ciliberto, M.~Herbster, A.~D. Ialongo, M.~Pontil, A.~Rocchetto, S.~Severini, and L.~Wossnig, ``Quantum machine learning: a classical perspective,'' \emph{Proceedings of the Royal Society A: Mathematical, Physical and Engineering Sciences}, vol. 474, no. 2209, p. 20170551, 2018.

\bibitem{ayerdi2020qos}
J.~Ayerdi, S.~Segura, A.~Arrieta, G.~Sagardui, and M.~Arratibel, ``{QoS}-aware metamorphic testing: An elevation case study,'' in \emph{2020 IEEE 31st International Symposium on Software Reliability Engineering (ISSRE)}, 2020, pp. 104--114.

\bibitem{gartziandia2022machine}
A.~Gartziandia, A.~Arrieta, J.~Ayerdi, M.~Illarramendi, A.~Agirre, G.~Sagardui, and M.~Arratibel, ``Machine learning-based test oracles for performance testing of cyber-physical systems: An industrial case study on elevators dispatching algorithms,'' \emph{Journal of Software: Evolution and Process}, vol.~34, no.~11, p. e2465, 2022.

\bibitem{oslocity}
Helseetaten, ``Norwegian health authority,'' \url{https://www.oslo.kommune.no/etater-foretak-og-ombud/helseetaten/}, 2024, [Online; accessed 20-Aug-2025].

\bibitem{sartaj2023hita}
H.~Sartaj, S.~Ali, T.~Yue, and J.~M. Gjøby, ``{HITA: An Architecture for System-level Testing of Healthcare IoT Applications},'' in \emph{European Conference on Software Architecture}.\hskip 1em plus 0.5em minus 0.4em\relax Cham: Springer, 2024, pp. 451--468.

\bibitem{sartaj2024modelbased}
H.~Sartaj, S.~Ali, T.~Yue, and K.~Moberg, ``Model-based digital twins of medicine dispensers for healthcare {IoT} applications,'' \emph{Software: Practice and Experience}, vol.~54, no.~6, pp. 1172--1192, 2024.

\bibitem{sartaj2025restapi}
H.~Sartaj, S.~Ali, and J.~M. Gjøby, ``{REST API Testing in DevOps: A Study on an Evolving Healthcare IoT Application},'' \emph{ACM Transactions on Software Engineering and Methodology}, pp. 1--45, 2025.

\bibitem{sartaj2024uncertainty}
------, ``Uncertainty-aware environment simulation of medical devices digital twins,'' \emph{Software and Systems Modeling}, vol.~24, no.~3, pp. 651--677, 2025.

\bibitem{arcuri2018evomaster}
A.~Arcuri, ``{EvoMaster}: Evolutionary multi-context automated system test generation,'' in \emph{2018 IEEE 11th International Conference on Software Testing, Verification and Validation (ICST)}.\hskip 1em plus 0.5em minus 0.4em\relax IEEE, 2018, pp. 394--397.

\bibitem{laaber2023challenges}
C.~Laaber, T.~Yue, S.~Ali, T.~Schwitalla, and J.~F. Nyg{\aa}rd, ``Challenges of testing an evolving cancer registration support system in practice,'' in \emph{Proceedings of the 45th {IEEE}/{ACM} International Conference on Software Engineering: Companion Proceedings}, ser. {ICSE}-Companion 2023.\hskip 1em plus 0.5em minus 0.4em\relax IEEE, 2023, pp. 355--359.

\bibitem{featureimportance}
Z.~Zhou and G.~Hooker, ``Unbiased measurement of feature importance in tree-based methods,'' \emph{ACM Trans. Knowl. Discov. Data}, vol.~15, no.~2, Jan. 2021.

\bibitem{qiskit}
A.~Javadi-Abhari, M.~Treinish, K.~Krsulich, C.~J. Wood, J.~Lishman, J.~Gacon, S.~Martiel, P.~D. Nation, L.~S. Bishop, A.~W. Cross, B.~R. Johnson, and J.~M. Gambetta, ``Quantum computing with {Qiskit},'' \emph{CoRR}, vol. abs/2405.08810, 2024.

\bibitem{quri}
\BIBentryALTinterwordspacing
QuanSys, ``{QURI Parts},'' 2024. [Online]. Available: \url{https://github.com/QunaSys/quri-parts}
\BIBentrySTDinterwordspacing

\bibitem{qraft}
T.~Patel and D.~Tiwari, ``Qraft: reverse your quantum circuit and know the correct program output,'' ser. ASPLOS '21.\hskip 1em plus 0.5em minus 0.4em\relax New York, NY, USA: Association for Computing Machinery, 2021, p. 443–455.

\bibitem{recomended_effectsize}
M.~Tomczak and E.~Tomczak, ``{The need to report effect size estimates revisited. An overview of some recommended measures of effect size},'' \emph{Trends in sport sciences}, vol.~21, no.~1, 2014.

\bibitem{statistics2}
A.~Arcuri and L.~Briand, ``{A Practical Guide for Using Statistical Tests to Assess Randomized Algorithms in Software Engineering},'' in \emph{Proceedings of the 33rd International Conference on Software Engineering}, ser. ICSE '11.\hskip 1em plus 0.5em minus 0.4em\relax New York, NY, USA: Association for Computing Machinery, 2011, pp. 1--10.

\bibitem{scoringrule}
\BIBentryALTinterwordspacing
Y.~Chung, W.~Neiswanger, I.~Char, and J.~Schneider, ``Beyond pinball loss: Quantile methods for calibrated uncertainty quantification,'' in \emph{Advances in Neural Information Processing Systems}, M.~Ranzato, A.~Beygelzimer, Y.~Dauphin, P.~Liang, and J.~W. Vaughan, Eds., vol.~34.\hskip 1em plus 0.5em minus 0.4em\relax Curran Associates, Inc., 2021, pp. 10\,971--10\,984. [Online]. Available: \url{https://proceedings.neurips.cc/paper_files/paper/2021/file/5b168fdba5ee5ea262cc2d4c0b457697-Paper.pdf}
\BIBentrySTDinterwordspacing

\bibitem{reliability}
H.~Gweon and H.~Yu, ``How reliable is your reliability diagram?'' \emph{Pattern Recognition Letters}, vol. 125, pp. 687--693, 2019.

\bibitem{barrier}
{\v{Z}}.~Vujovi{\'c} \emph{et~al.}, ``Classification model evaluation metrics,'' \emph{International Journal of Advanced Computer Science and Applications}, vol.~12, no.~6, pp. 599--606, 2021.

\bibitem{kitchenham2017robust}
B.~Kitchenham, L.~Madeyski, D.~Budgen, J.~Keung, P.~Brereton, S.~Charters, S.~Gibbs, and A.~Pohthong, ``Robust statistical methods for empirical software engineering,'' \emph{Empirical Softw. Engg.}, vol.~22, no.~2, pp. 579--630, apr 2017.

\bibitem{threats}
D.~S. Cruzes and L.~ben Othmane, ``Threats to validity in empirical software security research,'' in \emph{Empirical research for software security}.\hskip 1em plus 0.5em minus 0.4em\relax CRC Press, 2017, pp. 275--300.

\bibitem{graph}
A.~Bulu{\c{c}}, H.~Meyerhenke, I.~Safro, P.~Sanders, and C.~Schulz, \emph{Recent advances in graph partitioning}.\hskip 1em plus 0.5em minus 0.4em\relax Springer, 2016.

\bibitem{heiraricy}
S.~Bansal and M.~K. Rana, ``An efficient approach of regression testing using hierarchical decomposition slicing,'' \emph{International Journal of Advanced Research in Computer Science}, vol.~8, no.~7, 2017.

\bibitem{dataenc}
M.~B. Pande, ``A comprehensive review of data encoding techniques for quantum machine learning problems,'' in \emph{2024 Second International Conference on Emerging Trends in Information Technology and Engineering (ICETITE)}, 2024, pp. 1--7.

\bibitem{guerra2023probabilistic}
M.~Guerra, S.~Scardapane, and F.~M. Bianchi, ``Probabilistic load forecasting with reservoir computing,'' \emph{IEEE Access}, vol.~11, pp. 145\,989--146\,002, 2023.

\bibitem{domingo2024quantifying}
L.~Domingo, M.~Grande, F.~Borondo, and J.~Borondo, ``Quantifying the uncertainty of reservoir computing: Confidence intervals for time-series forecasting,'' \emph{Mathematics}, vol.~12, no.~19, p. 3078, 2024.

\bibitem{slrqml}
D.~Peral-García, J.~Cruz-Benito, and F.~J. García-Peñalvo, ``Systematic literature review: Quantum machine learning and its applications,'' \emph{Computer Science Review}, vol.~51, p. 100619, 2024.

\bibitem{quantOptForSESurvey2025}
M.~Zhang, Y.~Li, T.~Yue, and K.~Cai, ``Quantum optimization for software engineering: A survey,'' \emph{CoRR}, vol. abs/2410.15494, 2025.

\bibitem{WangRoadMap2025}
X.~Wang, S.~Ali, and P.~Arcaini, ``Quantum artificial intelligence for software engineering: the road ahead,'' \emph{CoRR}, vol. abs/2505.04797, 2025.

\bibitem{ZhaoQBSE2025}
J.~Zhao, ``Quantum-based software engineering,'' \emph{CoRR}, vol. abs/2505.23674, 2025.

\bibitem{bootqa}
X.~Wang, A.~Muqeet, T.~Yue, S.~Ali, and P.~Arcaini, ``Test case minimization with quantum annealers,'' \emph{ACM Trans. Softw. Eng. Methodol.}, vol.~34, no.~1, Dec. 2024.

\bibitem{bqtmizerIEEESoftware2025}
X.~Wang, S.~Ali, and P.~Arcaini, ``{BQTmizer}: A tool for test case minimization with quantum annealing,'' \emph{IEEE Software}, vol.~42, no.~5, pp. 51--57, 2025.

\bibitem{trovato2024reformulating}
\BIBentryALTinterwordspacing
A.~Trovato, M.~De~Stefano, F.~Pecorelli, D.~Di~Nucci, and A.~De~Lucia, ``Reformulating regression test suite optimization using quantum annealing - an empirical study,'' \emph{Int. J. Softw. Tools Technol. Transf.}, vol.~26, no.~6, pp. 767--780, 2024. [Online]. Available: \url{https://doi.org/10.1007/s10009-024-00775-w}
\BIBentrySTDinterwordspacing

\bibitem{qelm1}
W.~Xiong, G.~Facelli, M.~Sahebi, O.~Agnel, T.~Chotibut, S.~Thanasilp, and Z.~Holmes, ``On fundamental aspects of quantum extreme learning machines,'' \emph{CoRR}, vol. abs/2312.15124, 2023.

\bibitem{qelm3}
S.~K. Satpathy, V.~Vibhu, B.~K. Behera, S.~Al-Kuwari, S.~Mumtaz, and A.~Farouk, ``Analysis of quantum machine learning algorithms in noisy channels for classification tasks in the iot extreme environment,'' \emph{IEEE Internet of Things Journal}, vol.~11, no.~3, pp. 3840--3852, 2024.

\bibitem{he2025}
\BIBentryALTinterwordspacing
W.~He, Z.~Jiang, T.~Xiao, Z.~Xu, and Y.~Li, ``A survey on uncertainty quantification methods for deep learning,'' 2025. [Online]. Available: \url{https://arxiv.org/abs/2302.13425}
\BIBentrySTDinterwordspacing

\bibitem{gawlikowski2023survey}
J.~Gawlikowski, C.~R.~N. Tassi, M.~Ali, J.~Lee, M.~Humt, J.~Feng, A.~Kruspe, R.~Triebel, P.~Jung, R.~Roscher \emph{et~al.}, ``A survey of uncertainty in deep neural networks,'' \emph{Artificial Intelligence Review}, vol.~56, no. Suppl 1, pp. 1513--1589, 2023.

\bibitem{siddique2022survey}
T.~Siddique, M.~S. Mahmud, A.~M. Keesee, C.~M. Ngwira, and H.~Connor, ``A survey of uncertainty quantification in machine learning for space weather prediction,'' \emph{Geosciences}, vol.~12, no.~1, p.~27, 2022.

\bibitem{lopez2025}
\BIBentryALTinterwordspacing
L.~J.~L. López, S.~Elsharief, D.~A. Jorf, F.~Darwish, C.~Ma, and F.~E. Shamout, ``Uncertainty quantification for machine learning in healthcare: A survey,'' 2025. [Online]. Available: \url{https://arxiv.org/abs/2505.02874}
\BIBentrySTDinterwordspacing

\end{thebibliography}


\end{document}